\documentclass[reprint,amsmath,amssymb,aps,prstab]{revtex4-1}
\usepackage[english] {babel}
\usepackage{amsfonts,hyperref,graphicx,soul,natbib,mathrsfs}

\usepackage{epstopdf}
\usepackage{xcolor}

\newtheorem{theorem}{Theorem}

\newcommand\ld{\mathalpha{:}}

\newcommand\vio{\begin{bmatrix}

q_i^{0}\\

p_i^{0}\\
\end{bmatrix}}

\newcommand\dvi{\begin{bmatrix}

\dot{q}_i\\

\dot{p}_i\\
\end{bmatrix}}

\newcommand\vi{\begin{bmatrix}

q_i\\

p_i\\
\end{bmatrix}}

\newcommand\rvi{\begin{bmatrix}

0&& 1\\

-1 && 0\\
\end{bmatrix}}

\hypersetup{
  pdfauthor={S.S. Baturin}
}

\begin{document}

\title{Hamiltonian preserving nonlinear optics}

\author{S.S. Baturin}%
\email{s.s.baturin@gmail.com}%
\affiliation{Department of Electrical Engineering and Department of Physics, Northern Illinois University, DeKalb, IL 60115, USA}%
\date{\today}

\begin{abstract}
In this paper we present a method of constructing a nonlinear accelerator lattice that has an approximate integral of motion that is given upfront. The integral under consideration is a Hamiltonian in normalized (canonical) coordinates that is preserved by a lattice with a given accuracy. We establish a connection between the integrator of a Hamiltonian in normalized coordinates and a real lens arrangement. We apply known algorithms of high-order symplectic integrators, to produce several nonlinear lattices and show that this approach could improve the design of the nonlinear insert considered at the IOTA and UMER facilities. We also suggest new lattice design based on the Yoshida integrator.    
\end{abstract}

\maketitle
\section{Introduction}\label{sec:intr}

The concept of integrable and quasi-integrable nonlinear optics has recently attracted significant attention. Initially suggested by Danilov  (see Ref.\cite{Dan} and references therein) and refined by Danilov and Nagaitsev in Ref.\cite{DanNag}, the concept has been expanded to more realistic cases with space charge and chromaticity effects accounted for \cite{Webb,Wal}. Experimental demonstration of the integrable optics concept is currently being conducted at the IOTA facility at Fermilab \cite{IOTA,Ant_o} as well as at UMER ring at the University of Maryland \cite{Umer}. 

The main idea behind the integrable optics concept is a special insert of nonlinear magnets that is accommodated by a purely linear ring. The system is arranged in a way that the effective Hamiltonian for the lattice is almost time-independent and the potential produced by one nonlinear magnet warrants separation of variables, and thus a second integral of motion \cite{DanNag, chad_r}.  Initial designs of the nonlinear insert considered for both experiments \cite{Ant_o,Umer} were based on an idea of approximating the smooth nonlinear potential with a certain number of nonlinear magnets (17 in the case of IOTA and 7 in the case of UMER) with their strength scaled according to a prescription derived in Ref.\cite{DanNag} and placed equidistantly. Questions remain whether this number could be reduced further, and if the performance and design of the nonlinear insert could be further enhanced.

In this paper we introduce a general method of designing a nonlinear lattice based on known symplectic integration methods. After establishing a connection between the integrator of a smooth Hamiltonian in the normalized (canonical) coordinates, and a fragment of a real lattice, we demonstrate that the nonlinear lattice that preserves a given nonlinear smooth Hamiltonian could be implemented with only three nonlinear elements. As a representative example, we consider sextupole nonlinear inserts for the IOTA ring, octupole nonlinear inserts for the IOTA ring, and a toy-model FODO lattice with three nonlinear magnets introduced in Ref.\cite{DanNag}.           

The suggested method may be used in combination with other established tools like the normal form analysis \cite{Baz1,Baz2,TurchettiNOCE,Dragt,Forest1}, controls of symplectic maps \cite{HSM}, and methods for the increase of integrability \cite{CaryPRAB} to design lattices with high dynamic aperture.

\section{General approach}\label{sec:genap}

In this section we introduce notation, terminology and give a brief overview on symplectic integration of an autonomous Hamiltonian system. We slightly alter commonly used splitting of the flow to establish a connection between the integrator for a given Hamiltonian and a magnet arrangement of a real lattice. 

For the reader's convenience we list definitions of several terms that are used through the paper in the Appendix \ref{app:df}.

\label{S:2}

\subsection{Theoretical background}
In this section we will give a brief overview of the mathematical tools and terminology that we use in the paper. For more details on the numerical integration and advanced concepts we refer the reader to the original book \cite{GInt}.  Basic concepts about Lie algebras and geometric methods in differential equation theory could be found in Refs.~\cite{Arnold,Arnold2}, applications of this ideas to the accelerator physics problems could be found in for example Refs.~\cite{Dragt,Forest1}. 

Let us consider a nonlinear autonomous system in $\mathbb{R}^{2n}$
\begin{align}
\label{eq:in}
\dot{X}(t)&=f\left[X(t) \right], \\
X(0)&=X_0.
\end{align}
Here, $X=[q_1,p_1,q_2,p_2,...,q_n,p_n]^{\mathrm{T}}$ is a $2n$-dimensional vector of positions and momenta at a time $t$ and $X_0=[q_1^{0},p_1^{0},q_2^{0},p_2^{0},...,q_n^{0},p_n^{0}]^{\mathrm{T}}$ is the vector of initial conditions, $f(X)$ is a vector function called a \textit{vector field}. Here and throughout the paper, the dot above the letter denotes the full time derivative.  The \textit{flow}, $\phi_t$, of the system \eqref{eq:in} is the mapping that establishes a connection between initial condition $X_0$ and some point in time $\phi_t(X_0)=X(t)$.  

If the vector field $f(X)$ could be represented as  $f(X)=f_1(X)+f_2(X)$ such that systems   
\begin{align}
\dot{X}&=f_1\left(X\right), \\ \nonumber
\dot{X}&=f_2(X)
\end{align}
could be exactly integrated and the corresponding flows $\phi^{[1]}_t$ and $\phi^{[2]}_t$ could be explicitly found, then one can build an approximate flow $\Psi$ of the initial system as follows.

Let us consider a time step $h<1$. Then up to the order $\mathcal{O}(h^2)$ the approximate flow on a time mesh with the step $h$ is

\begin{align}
\label{eq:integ}
\Psi_{t=mh}=(\phi^{[1]}_h\circ\phi^{[2]}_h)^m~m\in\mathbb{N}.
\end{align} 
Mapping $\Psi_{t=h}$ is simply an integrator of the first order by $h$ (Euler integrator) of the system \eqref{eq:in}.
 
The possibility of such splitting, and even its symplectic (volume preserving) property for any autonomous Hamiltonian system, could be easily seen from the following considerations.

Let $\mathrm{H}$ be a Hamiltonian and $X$ still a $2n$ vector of coordinates and momenta. Then from the Hamiltonian equations, the trajectory of the system could be found from 
\begin{align}
\label{H:eq}
\dot{X}=-\ld\mathrm{H}\ld X.
\end{align}   
Here $\ld \mathrm{H}\ld $ is the Lie operator with an action on $g$ defined by the Poisson bracket of $g$ with $\mathrm{H}$ as \cite{Dragt}
\begin{align}
\ld \mathrm{H}\ld g\equiv\{\mathrm{H},g\}=\sum\limits_{i=1}^{n}\frac{\partial \mathrm{H}}{\partial q_i}\frac{\partial g}{\partial p_i}-\frac{\partial \mathrm{H}}{\partial p_i}\frac{\partial g}{\partial q_i}
\end{align}
The introduced Lie operator has a simple connection to a known Liouville operator $\ld \mathrm{H}\ld=-i\widehat{\mathbf{L}}$.

The solution to the equation \eqref{H:eq} could be represented as an exponent of the Lie operator acting on a vector of initial conditions (see for example \cite{Arnold})
\begin{align}
X(t)=\exp \left(-t\ld \mathrm{H} \ld\right)X_0.
\end{align} 
The operator exponent in the above expression, in general is an infinite series, however, for some special Hamiltonians this series naturally truncates at a finite number of terms leading to the exact expression for the map $\exp \left(-t\ld \mathrm{H} \ld\right)$.

Let us assume that Hamiltonian $\mathrm{H}$ could be split into $\mathrm{H}=\mathrm{H}_1+\mathrm{H}_2$ in a way that $\exp \left(-t\ld \mathrm{H}_1 \ld\right)$ and $\exp \left(-t\ld \mathrm{H}_2 \ld\right)$ could be evaluated explicitly.
As far as the Lie operators $\ld \mathrm{H}_1 \ld$ and $\ld \mathrm{H}_2 \ld$ do not commute, we utilize the  Baker-Campbell-Hausdorff (BCH) formula \cite{Haus} to evaluate the composition 
\begin{align}
&\exp \left(-t\ld \mathrm{H}_1 \ld\right)\circ\exp \left(-t\ld \mathrm{H}_2 \ld\right)=\nonumber \\ &\exp \left(-t\ld \mathrm{H}_1+\mathrm{H}_2\ld+\frac{t^2}{2}\ld\{\mathrm{H}_1,\mathrm{H}_2\}\ld-\mathcal{O}(t^3)\right).
\end{align}
If we introduce a time mesh with the step $h<1$ we will see that up to the order $\mathcal{O}(h^2)$, the composition of the exponents is the exponent of the sum of Lie operators,
\begin{align}
\label{eq:lc}
&\exp \left(-h\ld \mathrm{H}_1 \ld\right)\circ\exp \left(-h\ld \mathrm{H}_2 \ld\right)=\nonumber \\&\exp \left(-h\ld \mathrm{H}_1+\mathrm{H}_2\ld+\mathcal{O}(h^2)\right).
\end{align}
This means that the composition will preserve Hamiltonian $\mathrm{H}=\mathrm{H}_1+\mathrm{H}_2$ up to the order  $\mathcal{O}(h)$.

By comparing the approximate flow $\Psi_{t=h}$ Eq.\eqref{eq:integ} with Eq.\eqref{eq:lc} and establishing the connection $\exp \left(-h\ld \mathrm{H}_1 \ld\right)\equiv \phi^{[1]}_h$, $\exp \left(-h\ld \mathrm{H}_2 \ld\right)\equiv \phi^{[2]}_h$,  we conclude that splitting of the flows is closely connected with the possibility of splitting the Hamiltonian. The exponent of the Lie operator is a symplectic (volume preserving map) \cite{GInt,Arnold,Arnold2,Dragt}, the one-step integrator $\Psi_{t=h}$ is also symplectic as a composition of symplectic transformations is a symplectic transformation (group property).

The considerations above lead us to the following observation \cite{Tabor}: a system with continuous time could be approximated with a discrete system that preserves the smooth Hamiltonian up to a given level of accuracy. It is worth mentioning that in the case of a four dimensional phase space (transverse motion in accelerator is decoupled from longitudinal) when variables in the Hamiltonian could be separated, this leads to quasi-integrability \cite{DanNag}.

In the following sections, we study the case when the preserved Hamiltonian is close to the desired time independent one, and thus the dynamics are predefined by the desired Hamiltonian. We find magnet arrangements that have effective Hamiltonians of a predefined form, up to an error, that could be expressed as a power of the phase advance between nonlinear elements.    

\subsection{Discretization of the smooth Hamiltonian and high order integrators\label{ss2}}
The connection between a smooth Hamiltonian system and a map naturally arises from building an integrator that is essentially a discrete analog of a smooth system by definition \cite{Tabor,GInt}. The more accurate the integrator, the better it reproduces dynamics of the original system. In this section we give a brief derivation of the specific forms of the known integrators that we are going to utilize further.   

We consider a smooth Hamiltonian of the form 
\begin{align}
\label{eq:GH}
\mathrm{H}=\sum\limits_{i=1}^{n}\frac{q_i^2+p_i^2}{2}+V(q_1,q_2,...,q_n).
\end{align}
Here $V$ is the nonlinear potential of the form
\begin{align}
\label{eq:Vdef}
V(q_1,q_2,...,q_n)=\sum\limits_{j=3}^{l}a_j P_j(q_1,q_2,...q_n),
\end{align}
 $P_j$ is a homogeneous polynomial of the degree $j$, and $a_j$ is a constant.    

We split the Hamiltonian into a part that corresponds to  linear motion
\begin{align}
\label{eq:L}
\mathrm{H}_1=\sum\limits_{i=1}^{n}\frac{q_i^2+p_i^2}{2}
\end{align} 
and a part  
\begin{align}
\label{eq:N}
\mathrm{H}_2=V(q_1,q_2,...,q_n)
\end{align}
that combines all nonlinearities. 
According to Hamiltonian equations, the vector field (or simply the force) that corresponds to the Hamiltonian $\mathrm{H}_1$ is
\begin{align}
f_1(q_1,p_1,q_2,p_2,...,q_n,p_n)=\nonumber \\ [p_1,-q_1,p_2,-q_2,...,p_n,-q_n]^{\mathrm{T}}.
\end{align}
This leads to a system of $n$ independent pairs of equations
\begin{align}
\dvi=\rvi\vi,~i\in\mathbb{Z}(1,n). 
\end{align}
The flow of this system is an $n$ block-diagonal matrix $R_\psi$ of rotations with one block given as
\begin{align}
R^{(i)}_\psi=\begin{bmatrix}
\cos(\psi)&& \sin(\psi)\\
-\sin(\psi) && \cos(\psi)\\
\end{bmatrix}.
\end{align}
Solutions are independent pairs of $[q_i,p_i]^{\mathrm{T}}$ that are given in terms of the flow as
\begin{align}
\vi=R^{(i)}_\psi\vio. 
\end{align}
Here $[q_1^{0},p_1^{0},q_2^{0},p_2^{0},...,q_n^{0},p_n^{0}]^{\mathrm{T}}$ is a vector of initial conditions.

The vector field that corresponds to the Hamiltonian $\mathrm{H}_2$ is given by
\begin{align}
f_2(q_1,p_1,q_2,p_2,...,q_n,p_n)= \nonumber \\
\left[0,-\partial_{q_1}V,0,-\partial_{q_2}V,...,0,-\partial_{q_n}V\right]^{\mathrm{T}}. 
\end{align}
Here $\partial_{q_i}$ stands for the partial derivative by $q_i$. The corresponding system of differential equations again could be written as $n$ pairs of $[q_i,p_i]$, however, they are no longer completely independent.
 \begin{align}
\dvi=\begin{bmatrix}
0\\
-\partial_{q_i}V(q_1,q_2,...,q_n)\\
\end{bmatrix},~i\in\mathbb{Z}(1,n).
\end{align}
As far as the vector field keep $q_i$ unchanged and only modifies $p_i$, the flow of the system above is defined as
\begin{align}
K_\psi X_0=[q^{0}_1,p_1^{0}-\psi \partial_{q_1}V,~...~,q^{0}_n,p_n^{0}-\psi\partial_{q_n}V]^{\mathrm{T}},
\end{align}  
with each partial derivative taken at the initial point $(q^{0}_1,q^{0}_2,...q^{0}_n)$.

\begin {figure}[t]
 \centering
\includegraphics[scale=0.6]{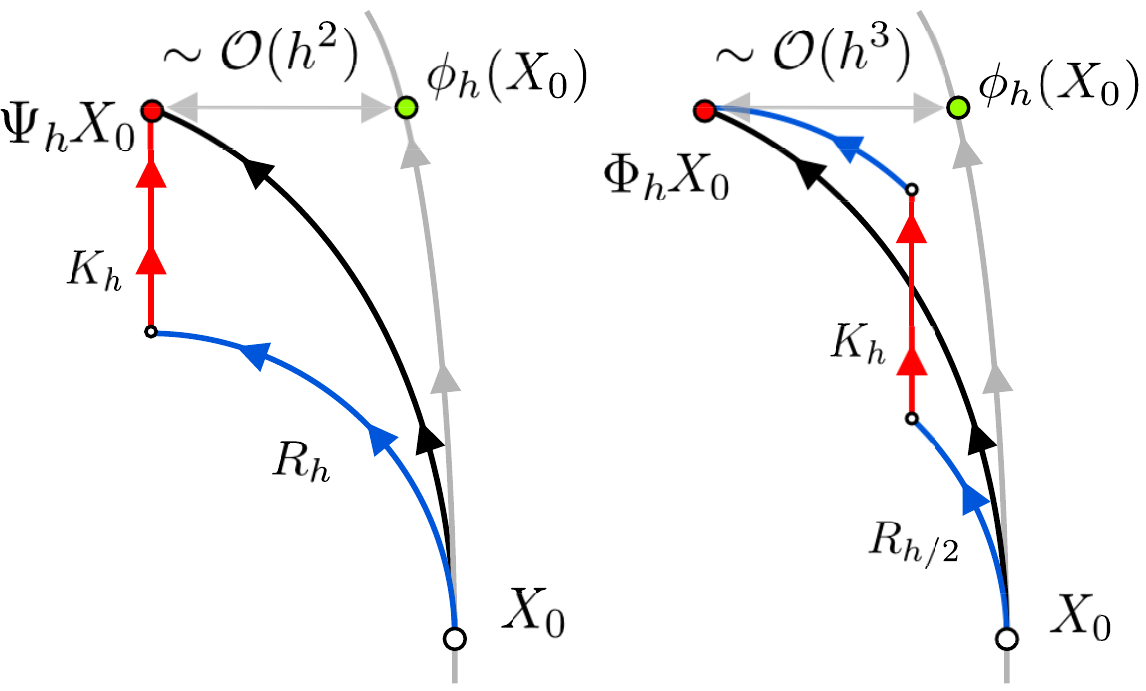}\\
\caption{Schematic diagrams: one step of the symplectic Euler method $\Psi_h$ (left) and one step of the second order Ruth method $\Phi_h$ (right). Grey line indicates exact flow $\phi_t$.}
\label{Fig:1}
\end {figure}

Now if we consider a time mesh, $t=mh~m\in\mathbb{N}$, with a step $h$, then the one step integrator $\Psi_h$ of the Hamiltonian $\mathrm{H}$ will have the from
\begin{align}
\label{eq:Eu}
\Psi_h=K_h\circ R_h.
\end{align}
This is a well known symplectic Euler method (Fig.\ref{Fig:1} left diagram). 
Next we consider a composition $R_{h/2}\circ K_h \circ R_{h/2}$. Keeping in mind that the exponent of the Lie operator is simply the flow: $\exp \left(-h\ld \mathrm{H}_1 \ld\right)\equiv R_h$, $\exp \left(-h\ld \mathrm{H}_2 \ld\right)\equiv K_h$ and using the BCH formula, we note that \cite{GInt}
\begin{align}
\label{eq:BCHr}
&R_{h/2}\circ K_h \circ R_{h/2}=\nonumber \\ &\exp \left(-\frac{h}{2}\ld \mathrm{H}_1 \ld\right) \circ \exp \left(-h\ld \mathrm{H}_2 \ld\right) \circ \exp \left(-\frac{h}{2}\ld \mathrm{H}_1 \ld\right)= \\
&=\exp \left(-h\ld \mathrm{H}_1+\mathrm{H}_2\ld+\mathcal{O}(h^3)\right). \nonumber
\end{align}
Due to the symmetry of the composition terms of the order $\mathcal{O}(h^2)$ cancel out $\frac{h^2\ld\{\mathrm{H}_1,\mathrm{H}_2\}\ld}{4}+\frac{h^2\ld\{\mathrm{H}_2,\mathrm{H}_1\}\ld}{4}=0$.

From the calculations above it is apparent that the integrator 
\begin{align}
\label{eq:Bm}
\Phi_h=R_{h/2}\circ K_h \circ R_{h/2}
\end{align}
preserves the Hamiltonian $\mathrm{H}$ up to the order $\mathcal{O}(h^2)$ and thus has higher accuracy than the simple Euler method. The integrator Eq.\eqref{eq:Bm} is known \cite{GInt} as Strang \cite{Strang} splitting or Marchuk splitting \cite{March} as well as the second order integrator introduced by Ruth \cite{Ruth1} (Fig.\ref{Fig:1} right diagram). 

Having an integrator $\phi$ of  given order $p$, it is often useful to build an integrator of higher order by  composing $\phi$ with itself.
The following theorem gives a general method of building such a composition \cite{GInt}
\begin{theorem}
\label{TH1}
Let $\phi_h$ be a one step integrator of the order $p$. If $\gamma_1+~...~+\gamma_s=1$ and $\gamma_1^{p+1}+~...~+\gamma_s^{p+1}=0$ then the composition 
\begin{align}
\phi_{\gamma_s h}\circ~...~\circ \phi_{\gamma_1 h}\nonumber
\end{align} 
is an integrator of order at least $p+1$.
\end{theorem}
An important consequence of the above theorem for $s=3$,  $\phi_{\gamma_3 h}\circ\phi_{\gamma_2 h}\circ \phi_{\gamma_1 h}$, is the three step Yoshida integrator \cite{Forest,Suzuki,Yoshida} with corresponding gammas given by 
\begin{align}
\gamma_1=\gamma_3=\frac{1}{2-2^{1/(p+1)}},~~~\gamma_2=-\frac{2^{1/(p+1)}}{2-2^{1/(p+1)}}. 
\end{align}

By composing the integrator \eqref{eq:Bm} ($p=2$) we arrive at the integrator of order $4$ in the form
\begin{align}
\label{eq:Yosh}
&\Phi^{Y}_h=\Phi_{\gamma_3h}\circ\Phi_{\gamma_2 h}\circ\Phi_{\gamma_1 h}, \\
&\gamma_1=\gamma_3=\frac{1}{2-2^{1/3}},~~\gamma_2=-\frac{2^{1/3}}{2-2^{1/3}}. \nonumber
\end{align}
By straightforward implementation of the BCH formula one can check that the integrator above preserves the Hamiltonian $\mathrm{H}=\mathrm{H}_1+\mathrm{H}_2$ up to  terms of  order $\mathcal{O}(h^4)$.
To avoid negative time  steps we use the identity $R_{2\pi}=\mathcal{I}$ and write the integrator \eqref{eq:Yosh} in the final form as
\begin{align}
\label{eq:Yoshf}
\Phi^{Y}_h=&R_{\gamma_1h/2}\circ K_{\gamma_1h}\circ R_{2\pi-\kappa_1 h/2} \nonumber\\ \circ &K_{-\kappa_2 h}\circ R_{2\pi-\kappa_1 h/2} \circ K_{\gamma_1 h} \circ R_{\gamma_1 h/2},  \nonumber\\
\gamma_1=&\frac{1}{2-2^{1/3}},~\kappa_1=\frac{2^{1/3}-1}{2-2^{1/3}}, ~\kappa_2=\frac{2^{1/3}}{2-2^{1/3}}. 
\end{align}
We reiterate one distinctive difference between the commonly used splitting of the Hamiltonian for numerical integration and the one we used above. Commonly, the Hamiltonian is split into a part that purely depends on momentum and a part that is purely dependent on the spatial coordinates. This results in a well known integration method that is sometimes referred as the ``drift-kick" method in the most simple Euler implementation. In our case, we separated the Hamiltonian into $\mathrm{H}_1$ - corresponding to linear motion and $\mathrm{H}_2$ - corresponding to a purely nonlinear ``kick" (the same way as in Ref.\cite{Bella}). This splitting allows us to establish a direct connection between the integrator of the Hamiltonian in the normalized coordinates and a transformation that corresponds to a set of optical elements.    

\subsection{Splitting of the nonlinear potential}

\begin {figure}[t]
 \centering
\includegraphics[scale=0.6]{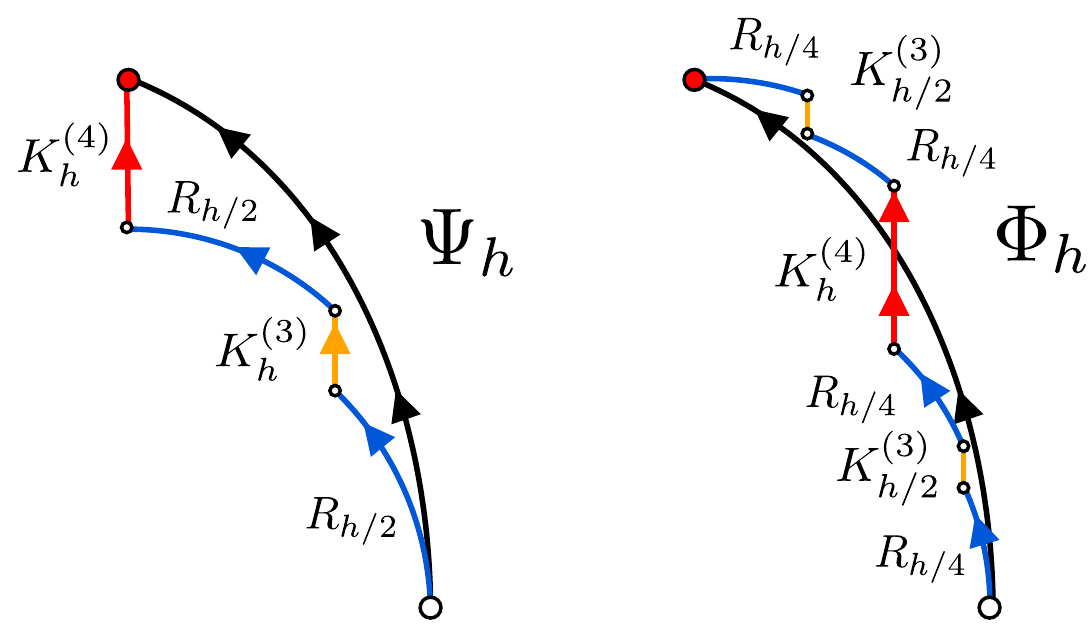}\\
\caption{Schematic diagrams of the one step of symplectic Euler method $\Psi_h$ (left) and second order Ruth method $\Phi_h$ (right) for the case of $l=4$ - nonlinear potential is split in to two parts (corresponding flows are $K^{(3)}$ and $K^{(4)}$).}
\label{Fig:2}
\end {figure}

We consider a nonlinear potential $V$ given by Eq.\eqref{eq:Vdef} and recall that it is essentially a sum of several potentials $V=\sum\limits_{j=3}^{l}a_j P_j$. Each one corresponds to a a specific order of the nonlinearity. As illustrated on Fig.\ref{Fig:2} we may rewrite the Euler method \eqref{eq:Eu}
in the form
\begin{align}
\label{eq:mNp}
\Psi_h=K_{h}^{(l)}\circ R_{h/(l-2)}\circ ...\circ K_{h}^{(3)}\circ R_{h/(l-2)}.
\end{align}  
Here $K_{h}^{(j)}$ is a flow that corresponds to the potential $a_j P_i$. Using the BCH formula one can ensure that this integrator indeed is of the order $\mathcal{O}(h)$ (preserves Hamiltonian $\mathrm{H_1}+\sum\limits_{j=3}^{l}a_j P_j$ up to the order $\mathcal{O}(h)$).

With a slight modification of the theorem \ref{TH1} one my show that the method 
\begin{align}
\label{eq:genR}
\Phi_h=\Psi_{h/2}^*\circ\Psi_{h/2}
\end{align}
with $\Psi_h^*=R_{h/(l-2)}\circ K_{h}^{(3)}\circ ...\circ R_{h/(l-2)}\circ K_{h}^{(l)}$ (adjoint method) is a method of the order 2 and thus preserves Hamiltonian $\mathrm{H_1}+\sum\limits_{j=3}^{l}a_j P_j$ up to the order $\mathcal{O}(h^2)$ (see Ref.\cite{GInt}).  
Further utilization of the method \eqref{eq:genR} by substituting it into the formula \eqref{eq:Yoshf} will result in increasing the order to the fourth order. 

Such splitting is useful for combining different types of nonlinear lenses in the same lattice.

\subsection{Connection to a real optical lattice}
We assume that the longitudinal and transverse motion are decoupled. Thus we consider a four dimensional phase space. We assume as well that there is no coupling in a linear lattice (transfer matrix has a block diagonal form).
  
To establish a connection between integrators in normalized coordinates $\{q_1,p_1,q_2,p_2\}$ and a real optical lattice, we recall that propagation of the particle from position $s_0$ to position $s_1$ through a linear optical channel could be described using a block diagonal transfer matrix \cite{SYL} with the block of the type
\begin{align}
\label{eq:ltr}
&M_{x,y}(s_1|s_0)=\nonumber \\& \mathbf{B}_{x,y}(s_1)
\begin{bmatrix}
\cos(\psi_{x,y})&& \sin(\psi_{x,y})\\
-\sin(\psi_{x,y}) && \cos(\psi_{x,y})\\
\end{bmatrix}
\mathbf{B}_{x,y}^{-1}(s_0).
\end{align} 
Here, the lower index denotes coordinate pair (either $\{x,P_x\}$ or $\{y,P_y\}$); $\psi_{x,y}=\int\limits_{s_0}^{s_1}\frac{ds}{\beta_{x,y}(s)}$ is the phase advance between position $s_0$ and $s_1$; $\mathbf{B}_{x,y}(s)$ is the corresponding block of the betatron amplitude matrix and $\mathbf{B}_{x,y}^{-1}(s)$ its inverse given by \cite{SYL}
\begin{align}
\label{eq:BTM}
\mathbf{B}_{x,y}(s)&=
\begin{bmatrix}
\sqrt{\beta_{x,y}(s)}&& 0\\
-\frac{\alpha_{x,y}(s)}{\sqrt{\beta_{x,y}(s)}} && \frac{1}{\sqrt{\beta_{x,y}(s)}}\\
\end{bmatrix}, \nonumber \\
\mathbf{B}_{x,y}^{-1}(s)&=
\begin{bmatrix}
\frac{1}{\sqrt{\beta_{x,y}(s)}}&& 0\\
\frac{\alpha_{x,y}(s)}{\sqrt{\beta_{x,y}(s)}} && \sqrt{\beta_{x,y}(s)} \\
\end{bmatrix}.
\end{align}
Here $\beta_{x,y}(s)$, and $\alpha_{x,y}(s)=-1/2\beta_{x,y}'(s)$ are the Twiss parameters of the linear lattice.

Now let us consider an integrator - $\Psi_h$, given by \eqref{eq:Bm} and propagate a vector of initial conditions for one step - $h$, that corresponds to the phase advance between the points $s_0$ and $s_2$, of a linear lattice 
\begin{align}
\label{eq:int}
X_h=R_{h/2}\circ K_{h}\circ R_{h/2} X_0.
\end{align}

With the identity $\mathcal{I}=\mathbf{B}(s)\circ\mathbf{B}^{-1}(s)$ (here $\mathcal{I}$ is the identity matrix) equation \eqref{eq:int} transforms as
\begin{align}
\mathrm{X}_h=M(s_2|s_1)\circ\mathbf{B}(s_1)\circ K_{h}\circ\mathbf{B}^{-1}(s_1) \circ M(s_1|s_0) \mathrm{X}_0.
\end{align} 
Here, $\mathrm{X}$  denotes the unnormalized state vector $\mathrm{X}_{h,0}\equiv\mathbf{B} X_{h,0}$.

We evaluate $\mathbf{B}(s_1)\circ K_{h}\circ\mathbf{B}^{-1}(s_1)$ further to achieve nonlinear element strength scaling with the $\beta$-function in a form
\begin{align}
\label{eq:nlt}
N^{\beta}\mathrm{X}_0=\left[x^0,P^0_x-h \frac{\partial_{x}U}{\sqrt{\beta_x}},y^0,P_y^{0}-h\frac{\partial_{y}U}{\sqrt{\beta_y}}\right],
\end{align} 
with 
\begin{align}
\label{eq:nlt2}
\partial_{x,y}U=\partial_{q_1,q_1}V(q_1,q_2)~q_1\to\frac{x}{\sqrt{\beta_x}},q_2\to\frac{y}{\sqrt{\beta_y}}.  
\end{align}

The two building blocks of the integrator are the flows $R_h$ and $K_h$, that now with the help of the Eq.\eqref{eq:ltr}, Eq.\eqref{eq:nlt} and Eq.\eqref{eq:nlt2} could be transformed to $M(s_2|s_1)$ and $N^{\beta}$ respectively. Maps $M(s_2|s_1)$ and $N^{\beta}$ could be implemented with thin lenses in a real lattice. We stress that the transformation for a nonlinear kick $N^{\beta}$ is now dependent on a $\beta$ function and thus has to be properly scaled to match the  linear part of the lattice. A similar result was achieved in \cite{DanNag}, however, the distribution of the magnets was considered to be continuous. The results of  Ref.\cite{DanNag} could be considered as a limit for the present approach, in the case of infinite number of nonlinear magnets (setting the method step $h\to 0$ to zero). For the reader's convenience the passage to the limit $h\to 0$ and recovery of the scaling derived in Ref.\cite{DanNag} is outlined in the Appendix \ref{app:DNlim}.      

\subsection{Remarks}

Having in hands all the necessary tools: the basic integration method that is given by Eq.\eqref{eq:Eu}, Theorem \ref{TH1} for increasing the order, splitting strategy given by Eq.\eqref{eq:mNp} and scaling of the nonlinear lens strength with respect to the linear optics from Eq.\eqref{eq:nlt} and Eq.\eqref{eq:nlt2}, one may design a nonlinear lattice that will upfront conserve any nonlinear Hamiltonian with a given accuracy. The splitting strategy in Eq.\eqref{eq:mNp}  allows to separate different nonlinear lenses in space, thus enabling a variety of lattices that may include sextupole, octupole and even higher order magnets at the same time if needed. 

An important observation that follows from  Sec.\ref{ss2} is that linear lattice configurations are not limited to the case of equal $\beta$-functions $\beta_x=\beta_y$, the $\beta$-function could differ as far as the condition $\psi_x=\psi_y$ is satisfied. This means that in order to be consistent with the integrator structure, only the  phase advance in $x$ and $y$ have to be equal.

If $J=q_1^2+p_1^2+q_2^2+p_2^2$ is the action and $a_j$ are the amplitudes of the nonlinearities as in Eq.
\eqref{eq:Vdef} then while action is less than unity $J<1$ and all $a_j<1$, higher orders in the BCH formula are suppressed in comparison to the lower orders, and the series could be thought of as a convergent series. Thus for the case $J<1$ and $a_j<1$, the whole scheme is stable as a first correction to the Hamiltonian that comes from the discretization error, is proportional to some positive power of action that is greater then $1$ and, on top of that, is multiplied by a small parameter $h^p$, where $p$ is the integrator order. 

Stability over many iterations (Hamiltonian preservation property) of the integrator, or equivalently, stability of the corresponding lattice over many revolutions, is guaranteed by the following theorem \cite{Ben,GInt} 

 \begin{theorem}
\label{TH2}
Let $\mathrm{H}$ (the Hamiltonian) be an analytic function $\mathrm{H}: D\to \mathbb{R}$ (where $D \subset R^{2d}$) and $\Phi_h(X)$ a symplectic numerical method of the order $p$ with the step size $h$. If the numerical solution generated by the integrator stays on the compact set $K \subset D$ then there exists a $h_0$ such that $\mathrm{H}(X_n)=\mathrm{H}(X_0)+\mathcal{O}(h^p)$ over exponentially long time intervals, $nh\leq e^{h_0/2h}$.
\end{theorem}
Here $d$ - is the number of degrees of freedom and $n$ - is the number of iterations.

Theorem \ref{TH2} essentially states that a bounded trajectory will remain bounded for exponentially long times and addresses potential concerns regarding long term stability of the corresponding lattice. Practically this means that bounded trajectories of the discrete system will be close to the trajectories of the corresponding smooth system for any realistic time (number of revolutions).

\begin {figure*}
 \centering
\includegraphics[scale=0.42]{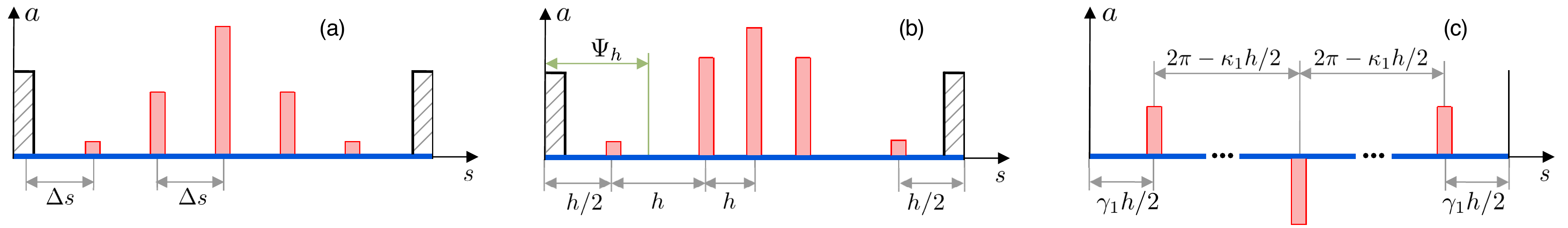}\\
\caption{Schematic diagrams of the nonlinear magnet layout for one period of the lattice: (a) equidistant placing as introduced in Ref.\cite{Ant_o}, (b) Ruth lattice, based on the Ruth second order integrator \eqref{eq:Bm} and Yoshida lattice (c) based on the Yoshida integrator \eqref{eq:Yoshf}. Here $s$ is the longitudinal spatial coordinate and $a$ is the normalized magnitude of the nonlinear magnets, $h$ is the phase advance between the magnets, multipliers $\gamma_1$ and $\kappa_1$ for Yoshida lattice are given by Eq\eqref{eq:Yoshf}.}
\label{Fig:3}
\end {figure*} 

Current consideration of an integrator based lattice has a tight connection to the problem of the resonance elimination and regularization of the particle motion, that is an established tool in creating nonlinear lattices with a large dynamic aperture (DA) \cite{CaryPRAB,TurchettiNOCE}.  Indeed the route to  large DA is to suppress higher order terms in the Lie exponent that could be constructed for example using the Birkhoff normal form approach \cite{TurchettiNOCE}. The key difference between prior art and present studies though, is in the direction of the analysis. We suggest to perform reverse engineering of a given smooth system and to find a lattice that has the best correspondence to this smooth system in terms of the dynamics. In contrast, in the common approach, derivation and analysis of the effective Hamiltonian for a given lattice is usually the focus. Interestingly, constructing a higher order integrator and a corresponding lattice we, according to the BCH formula, eliminate low order resonances and achieve the same goal. Thus we believe that the present idea may become a useful tool in developing nonlinear lattice design strategies.

\section{Examples of real lens configurations}\label{sec:examp}

In this section we consider two examples of the nonlinear lens arrangement that correspond to the Ruth second order integrator \eqref{eq:Bm} and Yoshida integrator \eqref{eq:Yoshf}. To provide head to head comparison in all cases of lattice tracking the integral strength of all nonlinear elements in the lattice was normalized to the unity
\begin{align}
\sum\limits_{i=1}^{k}a_0f\left(\beta(s_i))\right)=1
\end{align}
here $f(\beta(s))$ is the amplitude scaling function that is calculated with the help of the Eq.\eqref{eq:nlt} and Eq.\eqref{eq:nlt2}, $s_i$ is the position of the $i$-th nonlinear element and $a_0$ is a dimensional multiplier. The final total strength of the nonlinear channel were chosen such that the multiplier in-font of the nonlinear potential for the effective Hamiltonian is equal to unity. 

\subsection{Ruth lattice for the nonlinear insert}     

\begin {figure}[t]
 \centering
\includegraphics[scale=0.6]{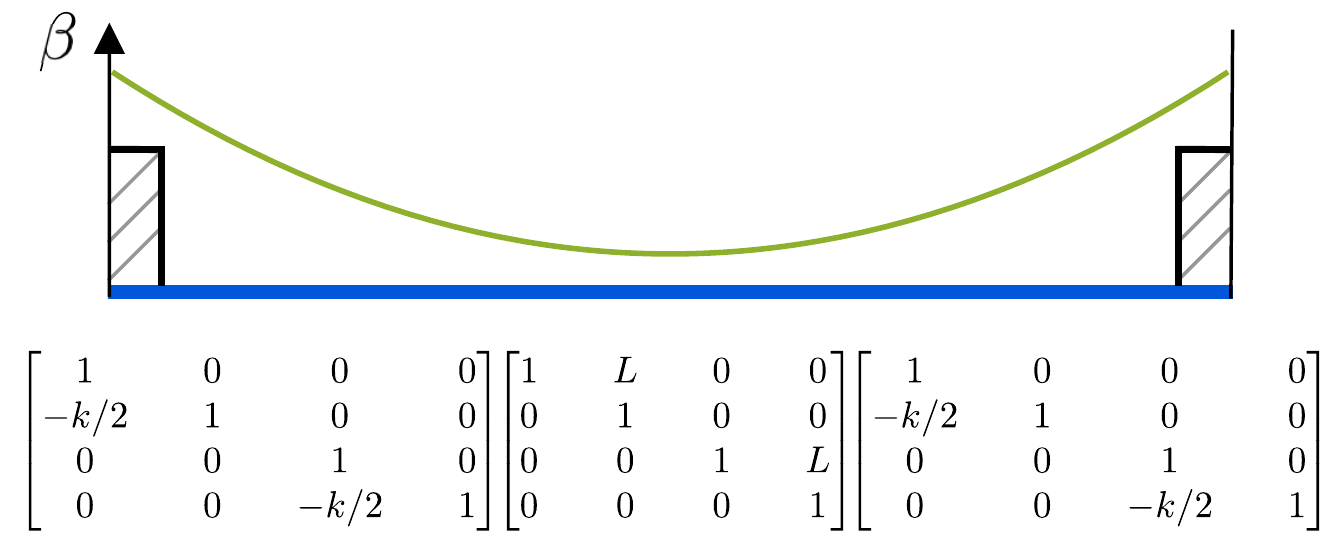}\\
\caption{Schematic diagram of  one period of the linear lattice with equal $\beta$-functions, introduced in Re.\cite{DanNag}. The whole ring is tuned to produce symmetric focusing in both $x$ and $y$ directions with the strength $k$ in the drift space of the length $L$.}
\label{Fig:4}
\end {figure}

We consider a lattice that is implemented in IOTA \cite{Ant_o} and UMER \cite{Umer} for the quasi-integrable and integrable optics experiments. The linear part of the lattice consists of the so called T-insert introduced in \cite{DanNag}, and a drift of length $L$. The T-insert is an arrangement where part of the linear optics effectively acts as a focusing matrix in both $x$ and $y$ directions, leading to a degenerate case of equal transverse $\beta$-functions in a drift space. This configuration is shown schematically in Fig.\ref{Fig:4}.

We define the total phase advance of the cell as $2\pi \nu$ where $\nu$ is the tune given by
\begin{align}
\nu=\frac{1}{2 \pi}\mathrm{arccos}\left(1-\frac{kL}{2} \right).
\end{align}
Here $k$ is the inverted focal length of the T-insert and $L$ is the length of the drift. The $\beta$-function is given by 
\begin{align}
\label{eq:bt}
\beta(s)=\frac{L-sk(L-s)}{\sqrt{1-\left(1-\frac{kL}{2}\right)}},
\end{align}
and the phase advance as a function of  position inside the drift
\begin{align}
\label{eq:pad}
\psi(s)=&\mathrm{arctan}\left(\sqrt\frac{kL}{4-kL}\right)-\nonumber \\&\mathrm{arctan}\left(\frac{\sqrt{k}L-2\sqrt{k}s}{\sqrt{L(4-kL)}}\right).
\end{align} 
As a reference, we will use the initial design of the nonlinear insert from Ref.\cite{Ant_o} where nonlinear magnets are placed with  equal distance in space and nonlinear potential scaling with respect to the $\beta$-function was chosen to be 
\begin{align}
\label{eq:DNp}
U(x,y,s)=\frac{1}{\beta(s)}V\left(\frac{x}{\sqrt{\beta(s)}},\frac{y}{\sqrt{\beta(s)}}\right).  
\end{align}
Schematically this arrangement is shown in Fig.\ref{Fig:3}(a).

From Eq.\eqref{eq:lc} and Eq.\eqref{eq:BCHr}, it is apparent that in order to maintain accuracy and integrator structure, it is required that nonlinear elements are separated by the same phase advance $h$.  Another important requirement is that $h<1$, as the error is proportional to the power of $h$. The linear stability criteria requires the tune, $\nu\leq0.5$ or equivalently the phase advance of the cell in Fig.\ref{Fig:4} to be less than $\pi$. In order to have a step $h<1$ in the whole range of tunes, the best choice of the number of integrator steps per one period is $N=5$. This gives a reasonable upper bound for the integrator step $h\leq\pi/5\approx 0.62$. 

As far as five phase steps are needed to fill the drift, we insert five nonlinear magnets according to the Ruth integration scheme of the second order with one step given by \eqref{eq:Bm}. With this lattice, the formula will read:
\begin{align}
\{T/2,O_1,N_1,O_2,N_2,O_3,N_3,O_3,N_2,O_2,N_1,O_1,T/2\} \nonumber
\end{align}
with each nonlinear magnet $N_j$ strength scaled to the $\beta$-function as prescribed by Eq.\eqref{eq:nlt} and Eq.\eqref{eq:nlt2}. Schematically this lattice is shown in Fig.\ref{Fig:3}(b). The length of each drift $O_j$ is calculated according to Eq.\eqref{eq:pad}. It is worth mentioning that in order to maintain the second order of the integrator, the first and the last drifts, $O_1$, in the lattice should correspond to half of the phase advance, $h/2$, between the nonlinear elements.    
\subsubsection{Sextupole channel\label{sec:rsex}}
For the first illustration of the Ruth lattice Fig.\ref{Fig:3}(b) based on the second order integrator, we consider the sextupole magnet as a nonlinear element. This type of nonlinear insert was fist proposed in Ref.\cite{AntHH} where authors suggested experiment for real world testing of regular and chaotic motion at IOTA ring based on Henon-Heiles system. 

The transverse part of the nonlinear potential of a thin sextupole is given by
\begin{align}
U^{(3)}(x,y)=\frac{a^{(3)}}{3}\left[3xy^2-x^3\right] , 
\end{align}
here $a^{(3)}$ is the strength of the sextupole. 
The corresponding smooth Hamiltonian for this system is a well studied Henon-Heiles Hamiltonian \cite{HenHal,Tabor}
\begin{align}
\label{eq:HH}
\mathrm{H}=\frac{p_1^2+p_2^2}{2}+\frac{q_1^2+q_2^2}{2}+q_1q_2^2-\frac{q_1^3}{3}.
\end{align}  
A system with this Hamiltonian is not integrable and exhibits chaotic motion. Nevertheless, trajectories  that correspond to the level set $\mathrm{H}>1/12$ may be  chaotic but they are still bounded up to the level set $\mathrm{H}=1/6$. This is a consequence of the fact that the equipotential lines are closed up to the level set $\mathrm{H}=1/6$ \cite{Tabor}.   

For comparison we build a lattice where sextupoles are placed with equal distance in accordance with the previous method of discretization from Ref.\cite{Ant_o} (Fig.\ref{Fig:3}(a)), 
as well as scale the sextupole strength with the $\beta$-function according to  Eq.\eqref{eq:DNp} following the Ref.\cite{DanNag} 
\begin{align}
\label{eq:DNpS}
a^{(3)}_{\mathrm{DN}}\sim \frac{1}{\beta^{5/2}(s_i)}.
\end{align}  
Here $s_i$ is the physical position of the $i$-th thin sextupole in the lattice.

In case of the Ruth lattice, nonlinear magnets are placed with equal phase advance between the magnets, with the first and last step being half of this phase advance. Positions in real space are calculated with Eq.\eqref{eq:pad}. The sextupole strength scaling with the $\beta$-function, with Eq.\eqref{eq:nlt}, Eq.\eqref{eq:nlt2} reads
\begin{align}
\label{eq:RP}
a^{(3)}_{\Psi}\sim \frac{1}{\beta^{3/2}(s_i)}.
\end{align}
Here, as before, $s_i$ - is the physical position of the $i$-th thin sextupole in the lattice.
We notice that the scaling law for the Ruth lattice is different from the original scaling \eqref{eq:DNpS} suggested in Ref.\cite{DanNag}.  

\begin {figure}[t]
 \centering
\includegraphics[scale=0.209]{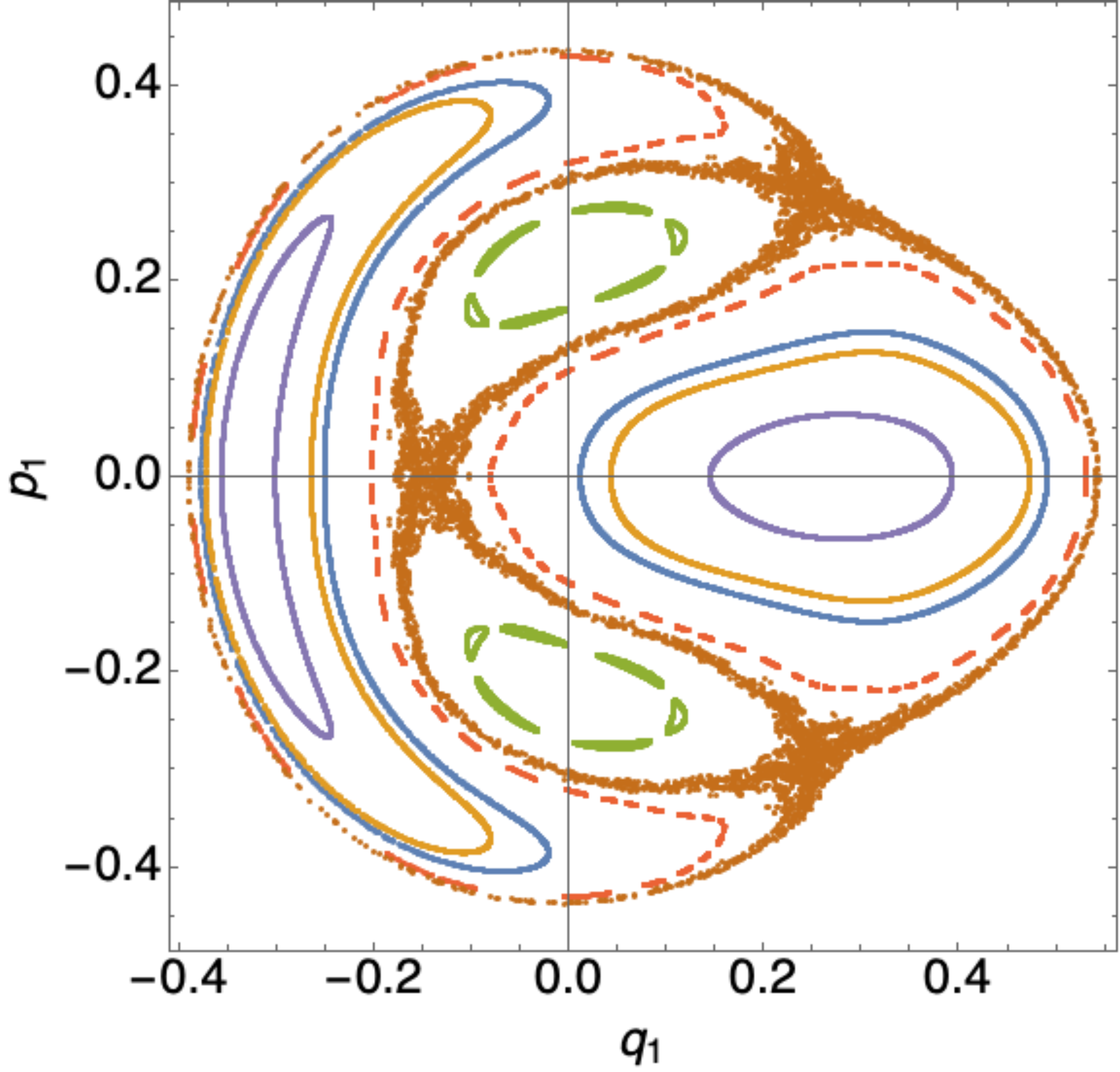}
\includegraphics[scale=0.209]{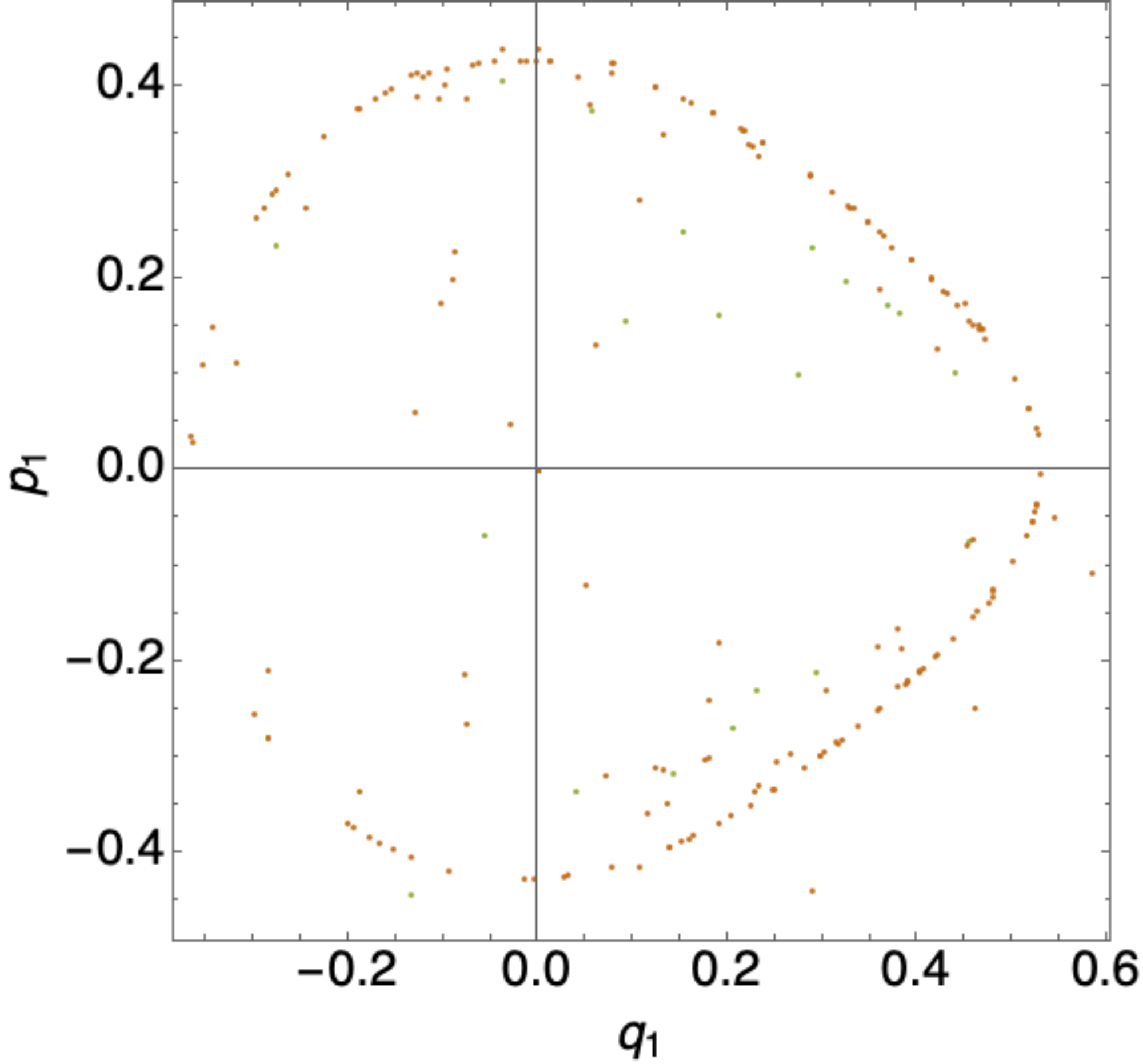}
\includegraphics[scale=0.43]{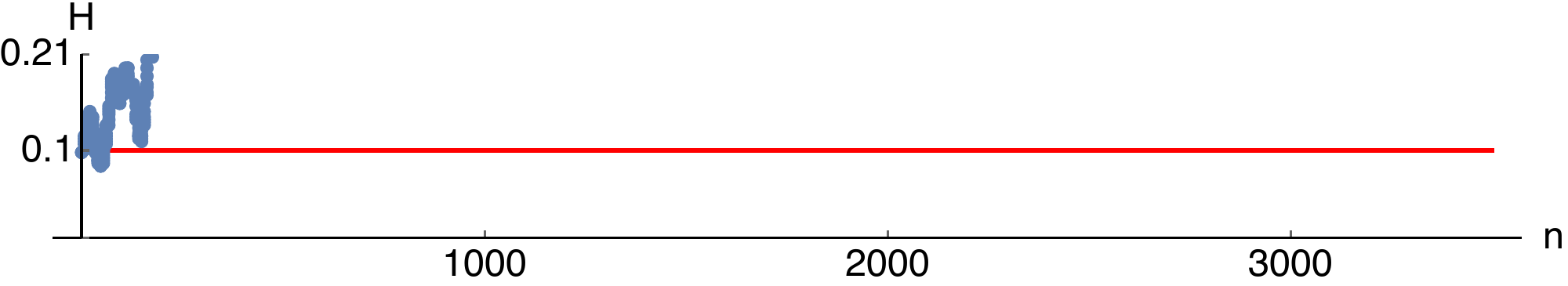}\\
\caption{Poincare surface of section $q_2=0$ for the smooth Henon-Heiles Hamiltonian Eq.\eqref{eq:HH} (left panel) and equidistant lattice introduced in Ref.\cite{Ant_o}  Fig.\ref{Fig:3}(a) with the sextupole magnet as a nonlinear element (right panel). Hamiltonian Eq.\eqref{eq:HH} as a function of the iteration number $n$ (lower panel). Blue line - tracking for the equidistant lattice and red line is the exact value of the Hamiltonian of the smooth system. We notice that the Hamiltonian Eq.\eqref{eq:HH} is not conserved and most of the initial conditions that were tracked escaped to infinity.}
\label{Fig:5e}
\end {figure}

To produce the comparison we chose the following parameters: Hamiltonian Eq.\eqref{eq:HH} level set $\mathrm{H}=0.1$ and the tune of the linear lattice $\nu=0.3344$ (slightly above the characteristic resonance $1/3$ of the sextupole).

First we build a Poincare surface of section $q_2=0$ for the equidistant lattice with the sextupole strength scaling given by \eqref{eq:DNpS} and compare it to the Poincare surface of a section of a smooth Hamiltonian in Fig.\ref{Fig:5e}.  We observe that tracking of the equidistant lattice failed as  most of the initial conditions rapidly escaped to infinity. Value of the Hamiltonian Eq.\eqref{eq:HH}  calculated from the tracking deviates significantly from the initial value of 0.1 and rapidly grows as it could be seen from the bottom panel in Fig.\ref{Fig:5e}.

\begin {figure}[t]
 \centering
\includegraphics[scale=0.209]{Fig5a.pdf}
\includegraphics[scale=0.209]{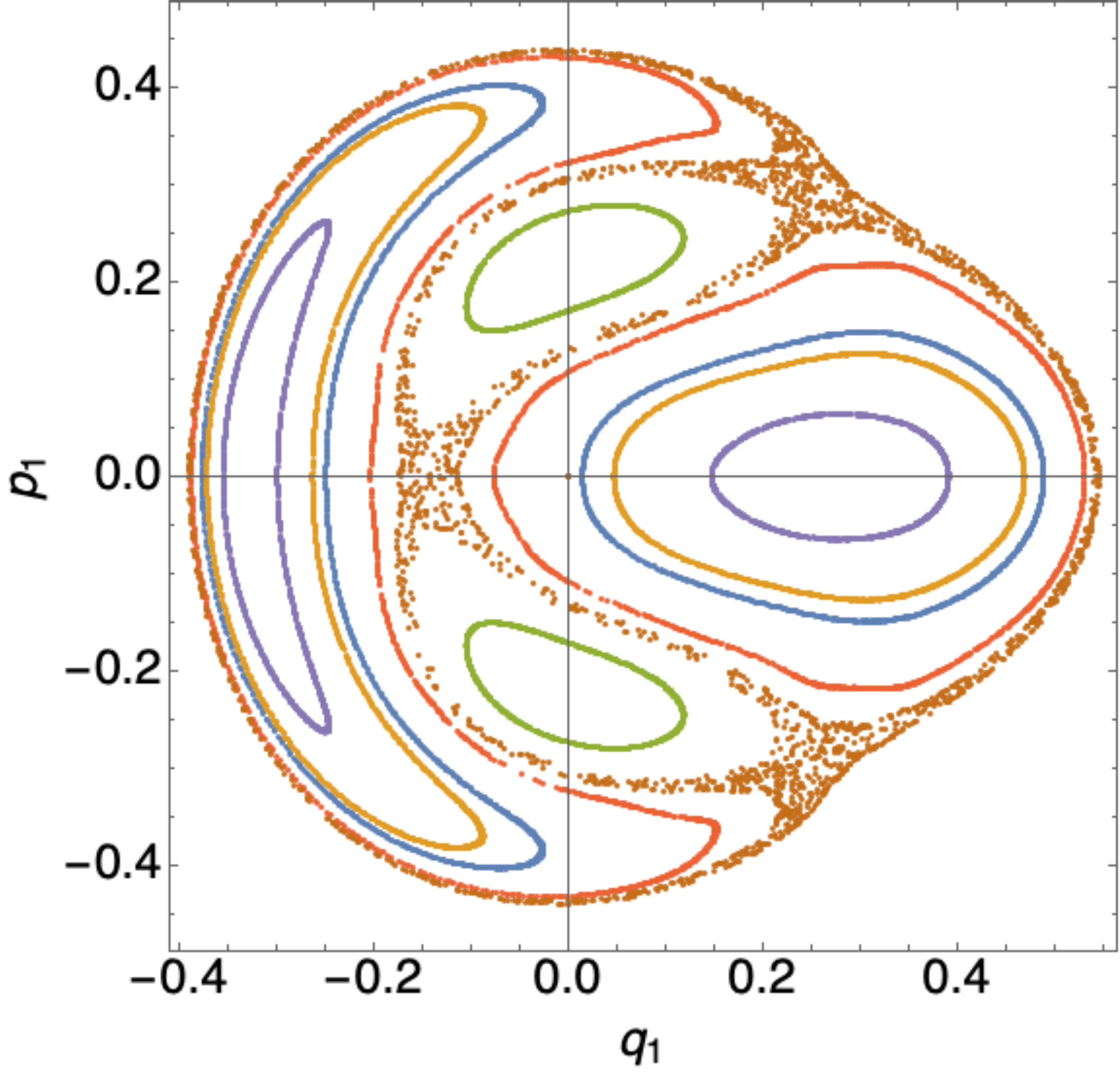}
\includegraphics[scale=0.43]{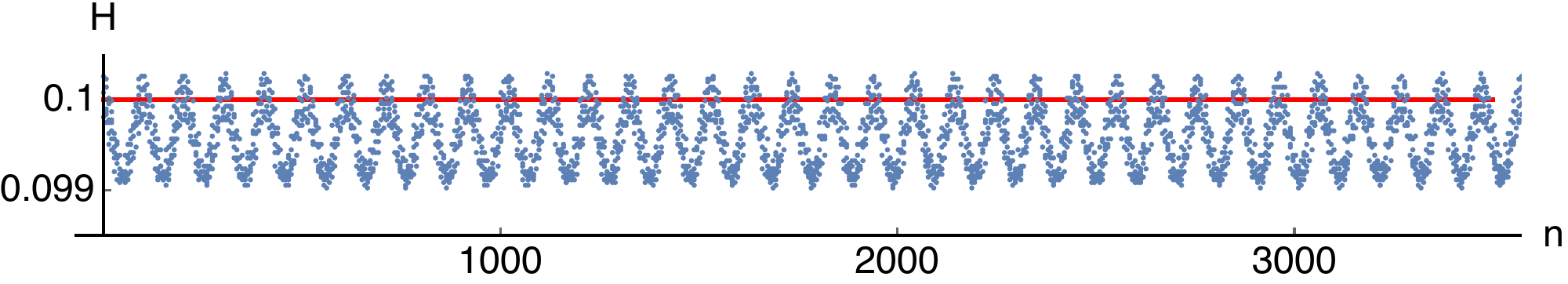}\\
\caption{Poincare surface of section $q_2=0$ for the smooth Henon-Heiles Hamiltonian Eq.\eqref{eq:HH} (left panel) and Ruth lattice with the sextupole magnet as a nonlinear element (right panel). Hamiltonian  Eq.\eqref{eq:HH} as a function of the iteration number $n$ (lower panel). Blue line - tracking for the Ruth lattice and red line is the value of the Hamiltonian of the smooth system. We notice that the Hamiltonian is conserved within an error of $\approx 1\% .$}
\label{Fig:5}
\end {figure}

Next we demonstrate that the Ruth lattice indeed preserves the Hamiltonian. We build a Poincare surface of section $q_2=0$ for the Ruth lattice and compare it to the Poincare surface of a section of a smooth Hamiltonian in Fig.\ref{Fig:5}. From the upper plots in Fig.\ref{Fig:5} it is apparent that the topology of both surfaces of the section coincide quite well and Hamiltonian (Fig.\ref{Fig:5} bottom panel) is conserved with high accuracy.

To produce Poincare surface of section in both cases (Ruth and equidistant lattice) several sets of initial conditions were tracked for $10^6$ iterations. Every point that was closer than $10^{-3}$ to the surface of section was projected on this surface. To produce Poincare surface of section of a smooth Hamiltonian 6th order symplectic integrator was used. Once the phase trajectory passed though the surface, a step back was made, and integration step was decreased. The procedure was repeated until the distance from the closet point to the surface of section was below $10^{-12}$. This point was considered to be located on the surface of section.

\begin {figure}[t]
 \centering
\includegraphics[scale=0.29]{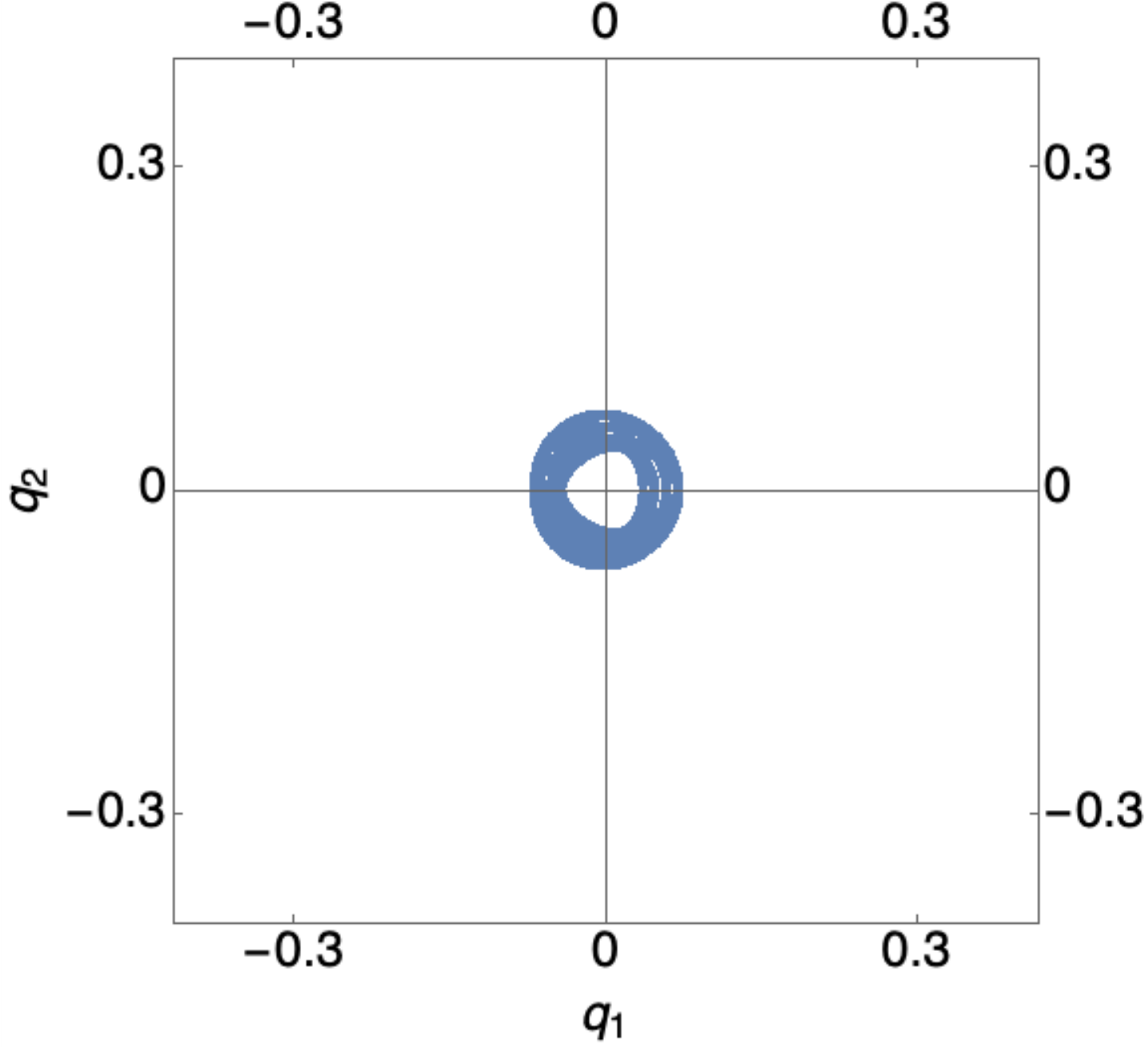}
\includegraphics[scale=0.29]{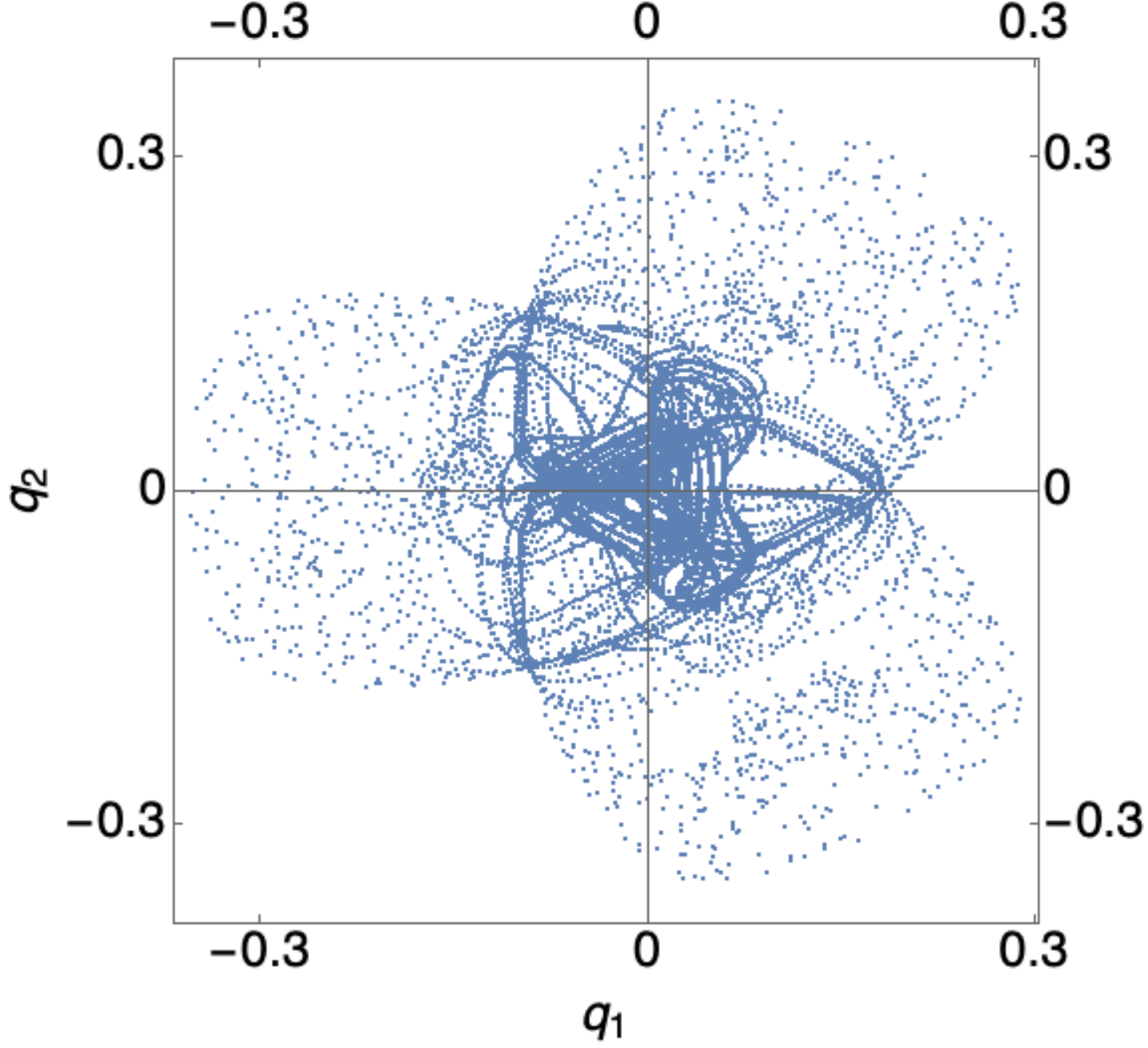}
\includegraphics[scale=0.29]{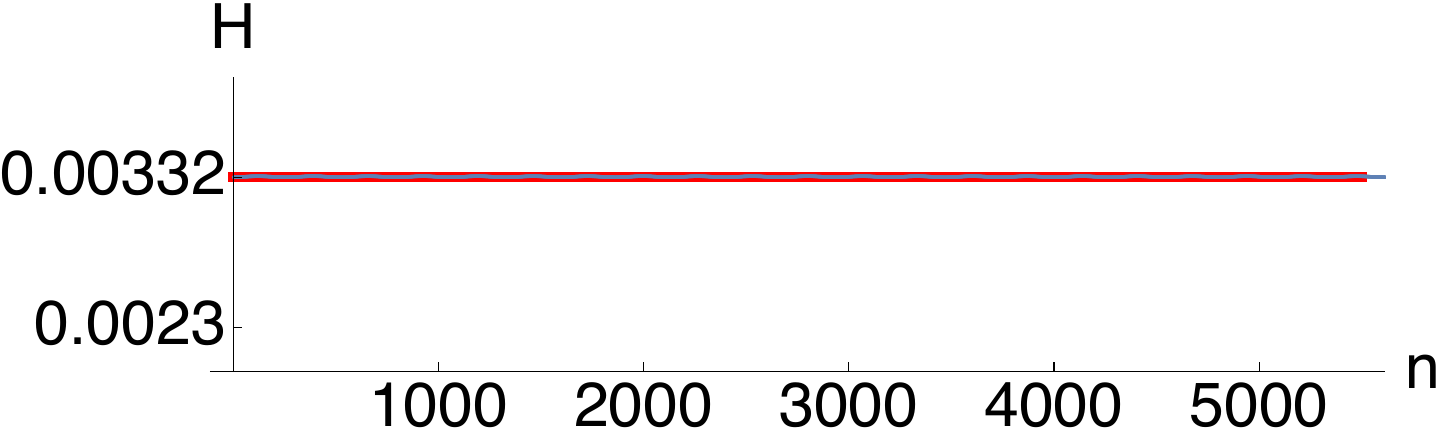}
\includegraphics[scale=0.29]{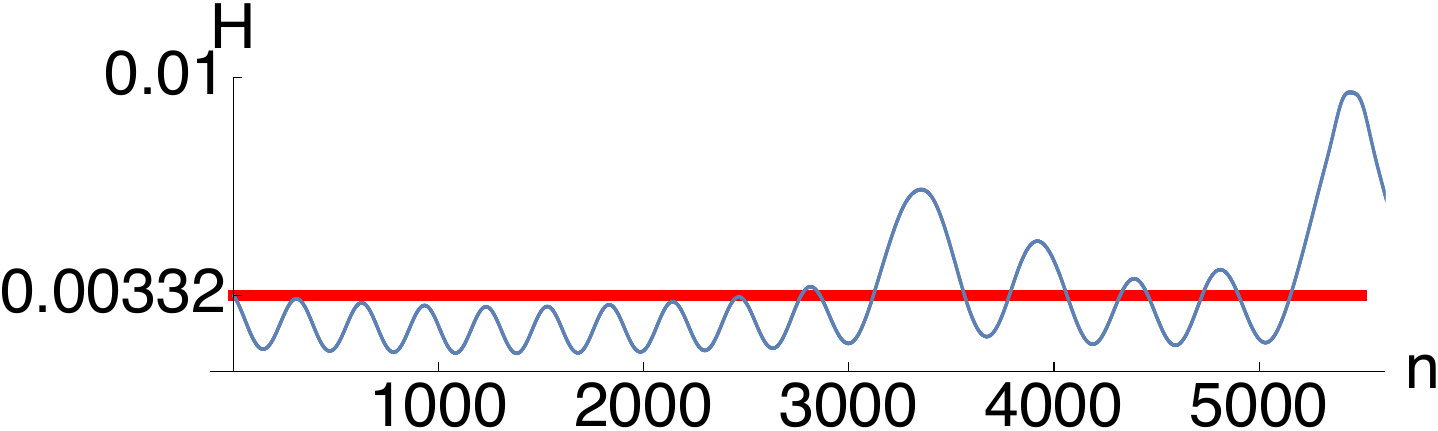}\\
\caption{Projection of the trajectory on the $(q_1,q_2)$ plane for the equidistant lattice with the sextupole magnet as a nonlinear element (upper right panel) and Ruth lattice with the sextupole magnet as a nonlinear element (upper left panel). Hamiltonian  Eq.\eqref{eq:HH} as a function of the iteration number $n$ for the equidistant lattice (lower right panel) and Ruth lattice (lower left panel). Blue line - is tracking and red line is the exact value of the Hamiltonian Eq.\ref{eq:HH}. Initial conditions for the tracking are $q_1=-0.04$,  $p_1=0.01$, $q_2=0$, $p_2=0.07$.}
\label{Fig:6}
\end {figure}

It is worth to mention that Poincare surfaces of section for a smooth system and for the Ruth lattice are not identical. Slight difference that is observed in the upper right and upper left plots Fig.\ref{Fig:5} are the result of a high value of the parameter $h\approx0.42$ - equivalent integrator step in phase for the Ruth lattice. From Fig.\ref{Fig:5} it is clearly seen that mean value of the effective Hamiltonian in case of a lattice is lower and this results in a different mean phase space trajectory that would closer correspond to the phase trajectory of a smooth system at a lower level set.

In order to produce another head to head comparison of the Ruth lattice and the equidistant lattice, we pick a much lower level set of the Hamiltonian Eq.\eqref{eq:HH} $\mathrm{H}=0.00332$ and track both lattices for $1.5\times10^4$ iterations. We again picked the tune of the linear lattice $\nu=0.3344$ slightly above the characteristic resonance $1/3$ of the sextupole for both Ruth lattice and equidistant lattice. Projection of the trajectory on the $(q_1,q_2)$ plane is presented in Fig.\ref{Fig:6} for both cases. We observe that in case of the equidistant lattice the projection looks irregular and spread in comparison to the Ruth lattice. We also observe that while the Ruth lattice preserves energy as expected (Fig.\ref{Fig:6} bottom left panel), for the equidistant lattice fluctuations in Hamiltonian are extremely high (Fig.\ref{Fig:6} bottom right panel).

\subsubsection{Octupole channel\label{sec:roct}}
Next, we move to a more practical application and consider an octupole as the  nonlinear element. It is stated in Ref.\cite{Ant_o} that implementation of an octupole channel with one invariant of motion could provide large betatron frequency spread, and thus paves the way to overcome fast coherent beam instabilities. As a suitable invariant, per the initial suggestion in Ref.\cite{DanNag}, for both experiments described in Ref.\cite{Ant_o} and in Ref.\cite{Umer}, the Hamiltonian in normalized coordinates was considered. 

The transverse part of the potential for a thin octupole is given by
\begin{align}
U^{(4)}(x,y)=\frac{a^{(4)}}{4}\left[x^4+y^4-6x^2y^2\right].
\end{align}
The corresponding smooth system is again a Henon-Heiles type system with the Hamiltonian
\begin{align}
\label{eq:HH2}
\mathrm{H}=\frac{p_1^2+p_2^2}{2}+\frac{q_1^2+q_2^2}{2}+\frac{q_1^4+q_2^4}{4}-\frac{3}{2}q_1^2q_2^2.
\end{align}  
Using the same reasoning as for the original Henon-Heiles system, one may show that due to the fact that all equipotential lines are closed up to the level set $\mathrm{H}=1/4$ motion remain bounded, however as before it may be also chaotic.

\begin {figure}[t]
 \centering
\includegraphics[scale=0.209]{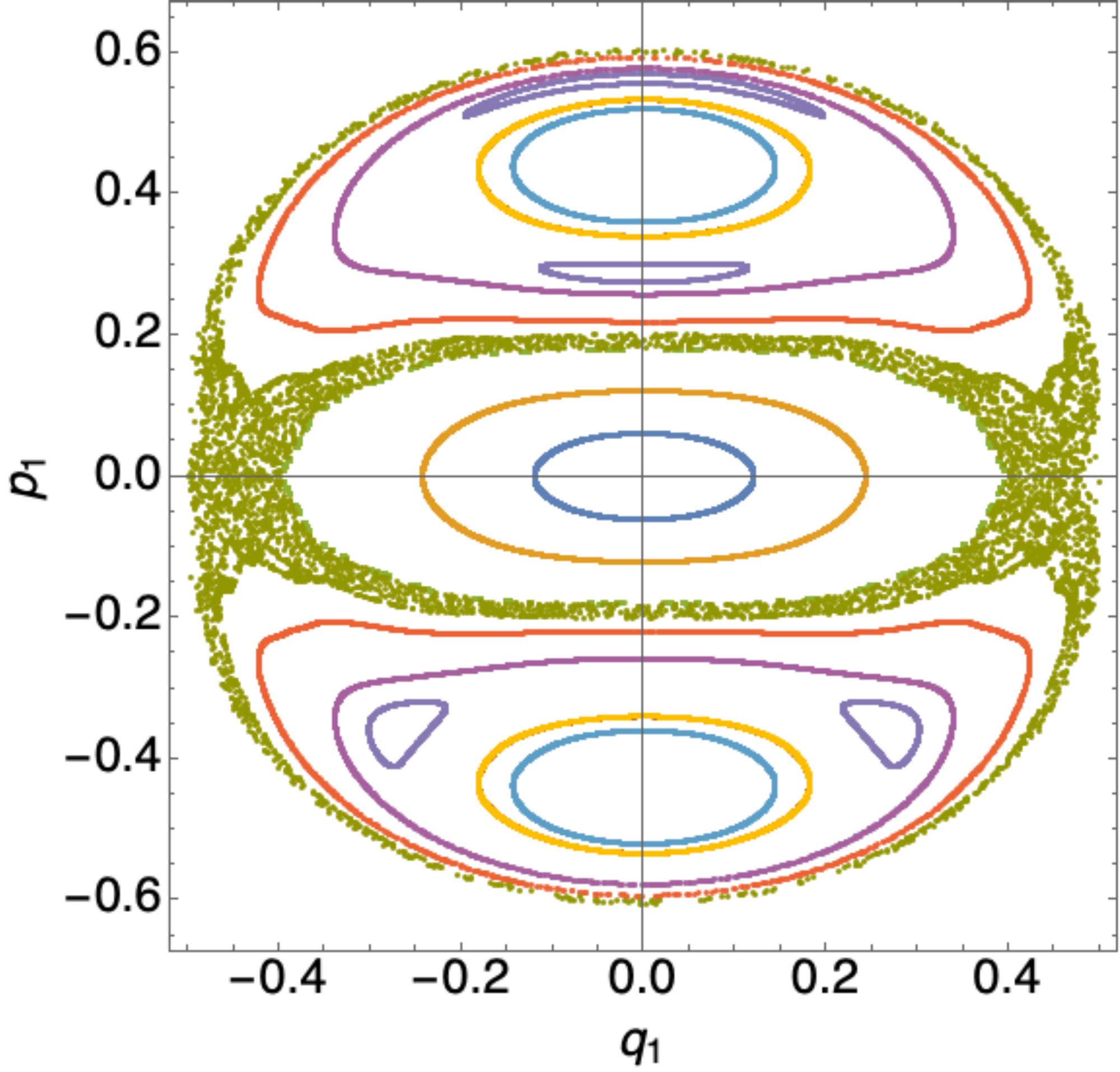}
\includegraphics[scale=0.209]{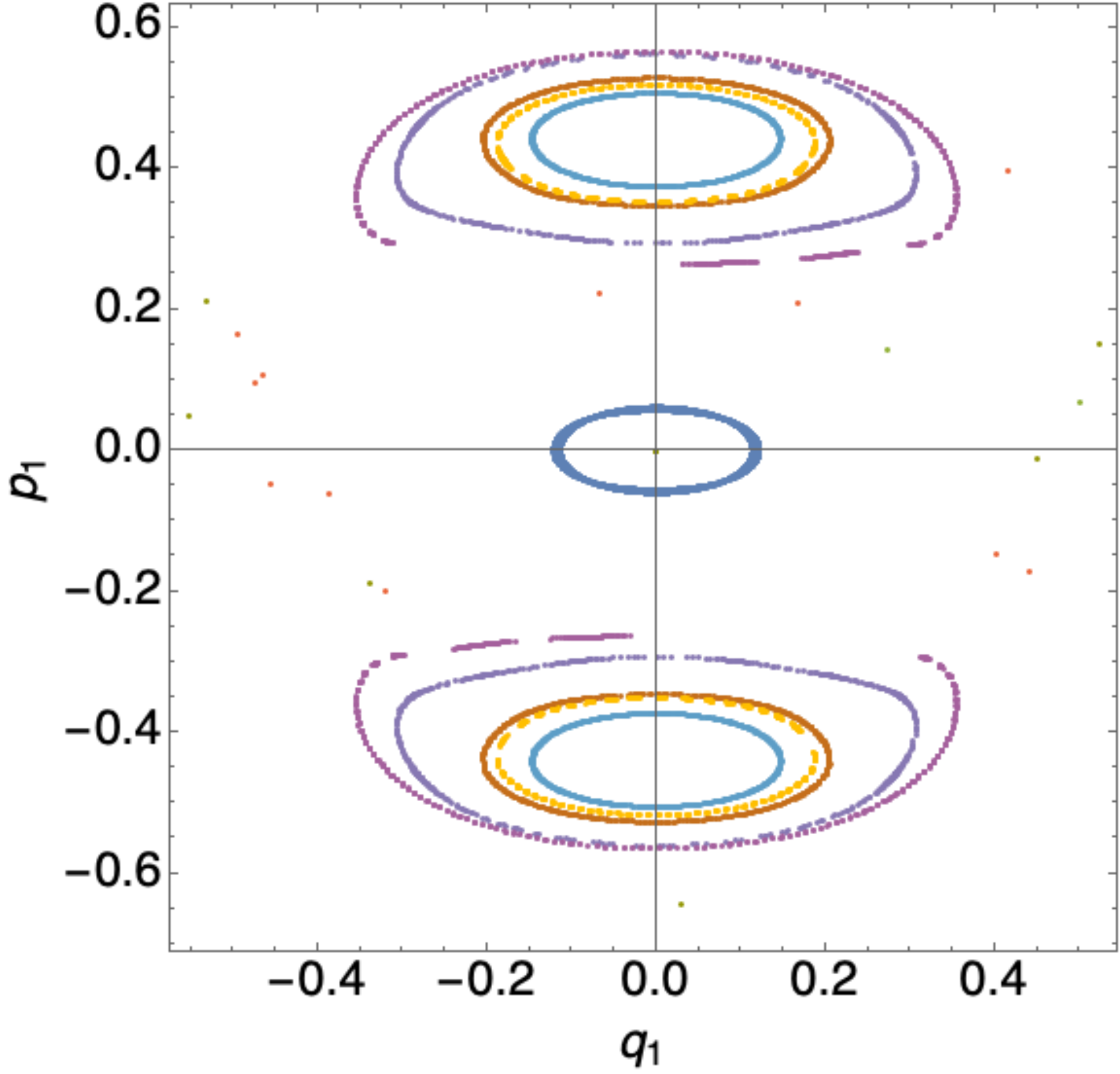}
\includegraphics[scale=0.43]{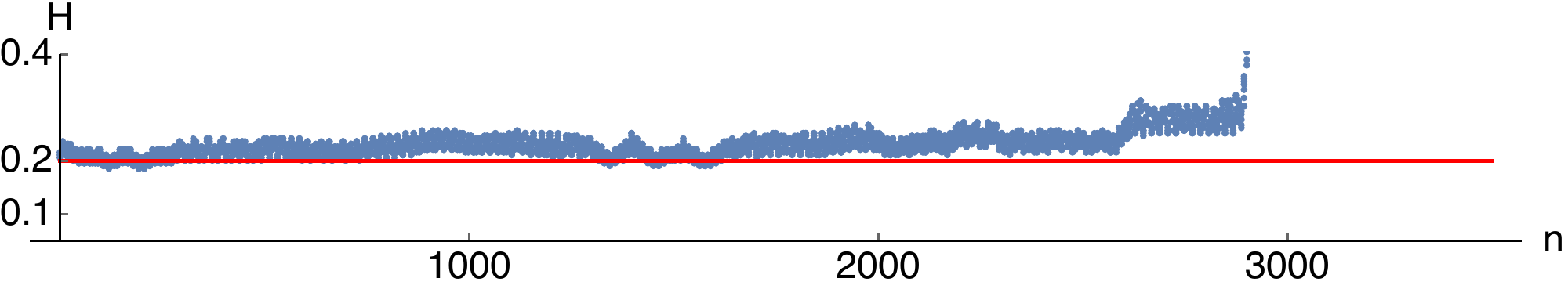}\\
\caption{Poincare surface of section $q_2=0$ for the smooth Henon-Heiles Hamiltonian Eq.\eqref{eq:HH2} (left panel) and equidistant lattice introduced in Ref.\cite{Ant_o}  (Fig.\ref{Fig:3}a) with the octupole magnet as a nonlinear element (right panel). Hamiltonian  Eq.\eqref{eq:HH2} as a function of the iteration number $n$ (lower panel). Blue line - tracking for equidistant lattice and red line is the exact value of the Hamiltonian of the smooth system. We notice that the Hamiltonian Eq.\eqref{eq:HH2} is conserved poorly in the beginning and after $\sim2800$ iterations completely deviate from the initial value of $0.2$.}
\label{Fig:7e}
\end {figure}  
   
According to the prescription of Ref.\cite{DanNag,Ant_o}, the octupole strength now should be scaled with the $\beta$-function according to Eq.\eqref{eq:DNp} as
\begin{align}
\label{eq:DNpO}
a_{\mathrm{DN}}^{(4)}\sim \frac{1}{\beta^3(s_i)},
\end{align}  
and again placed with equal distance between the magnets. Here $s_i$ - is the physical position of the $i$-th thin octupole in the lattice.
In the case of the Ruth lattice, octupole strength scaled with the $\beta$-function according to Eq.\eqref{eq:nlt} and Eq.\eqref{eq:nlt2} as
\begin{align}
a_{\Psi}^{(4)}\sim\frac{1}{\beta^2(s_i)}.
\end{align}
We notice that the scaling law for the Ruth lattice is different and has the same trend as in Eq.\eqref{eq:RP}: the power of $\beta(s)$ is less by $1$ in comparison to the Ref.\cite{DanNag}.

\begin {figure}[t]
 \centering
\includegraphics[scale=0.209]{Fig7a.pdf}
\includegraphics[scale=0.209]{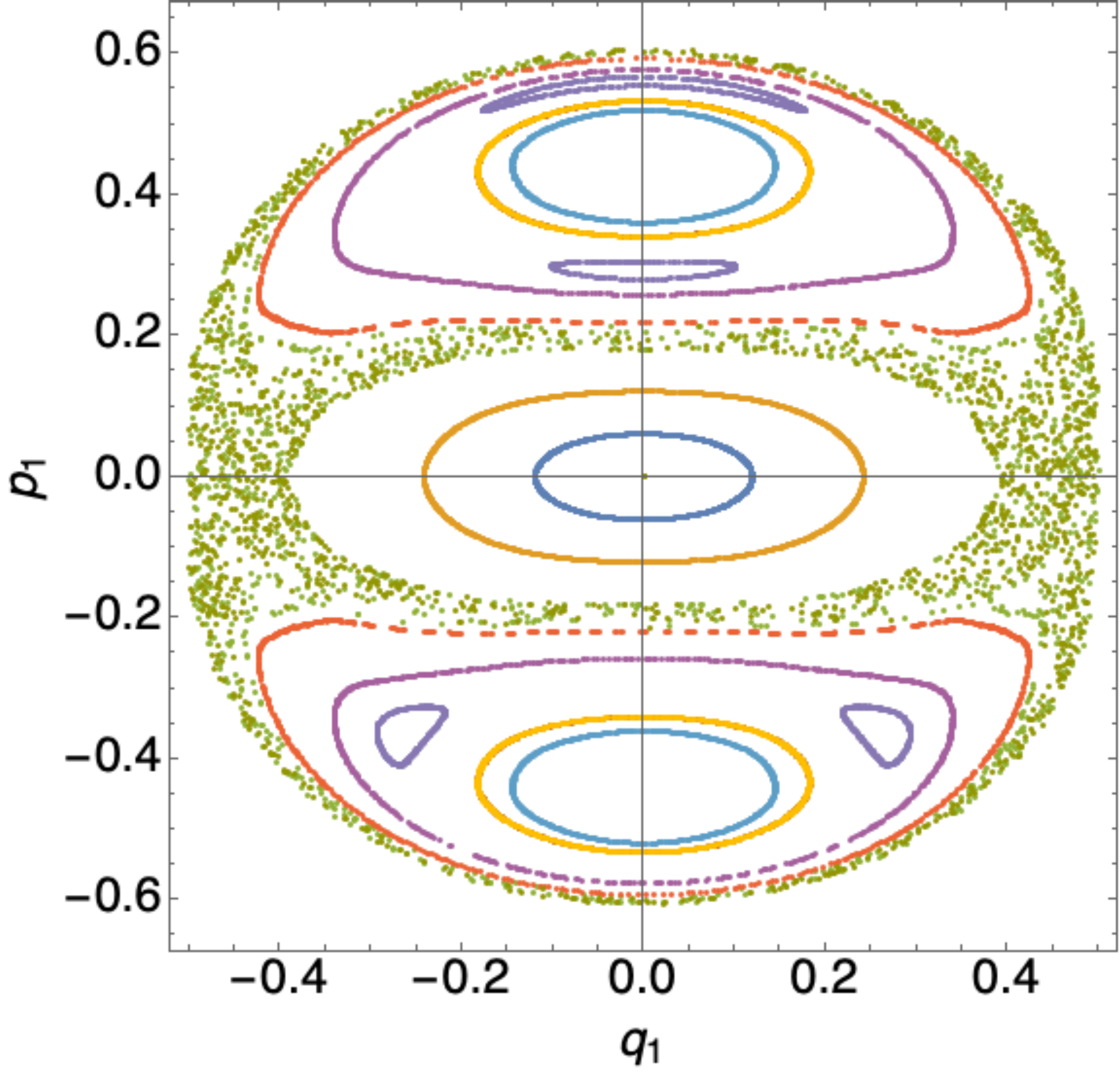}
\includegraphics[scale=0.43]{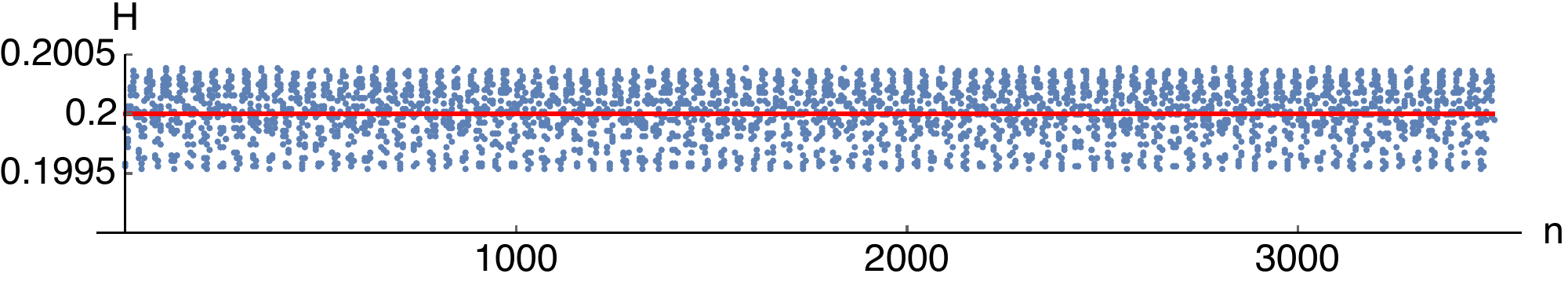}\\
\caption{Poincare surface of section $q_2=0$ for the smooth Henon-Heiles Hamiltonian Eq.\eqref{eq:HH2} (left panel) and Ruth lattice with the octupole magnet as a nonlinear element (right panel). Hamiltonian  Eq.\eqref{eq:HH2} as a function of the iteration number $n$ (lower panel). Blue line - tracking for the Ruth lattice and red line is the exact energy of the smooth system. We notice that the Hamiltonian is conserved within an error of $\approx 1\%.$}
\label{Fig:7}
\end {figure}      

\begin {figure}[t]
 \centering
\includegraphics[scale=0.29]{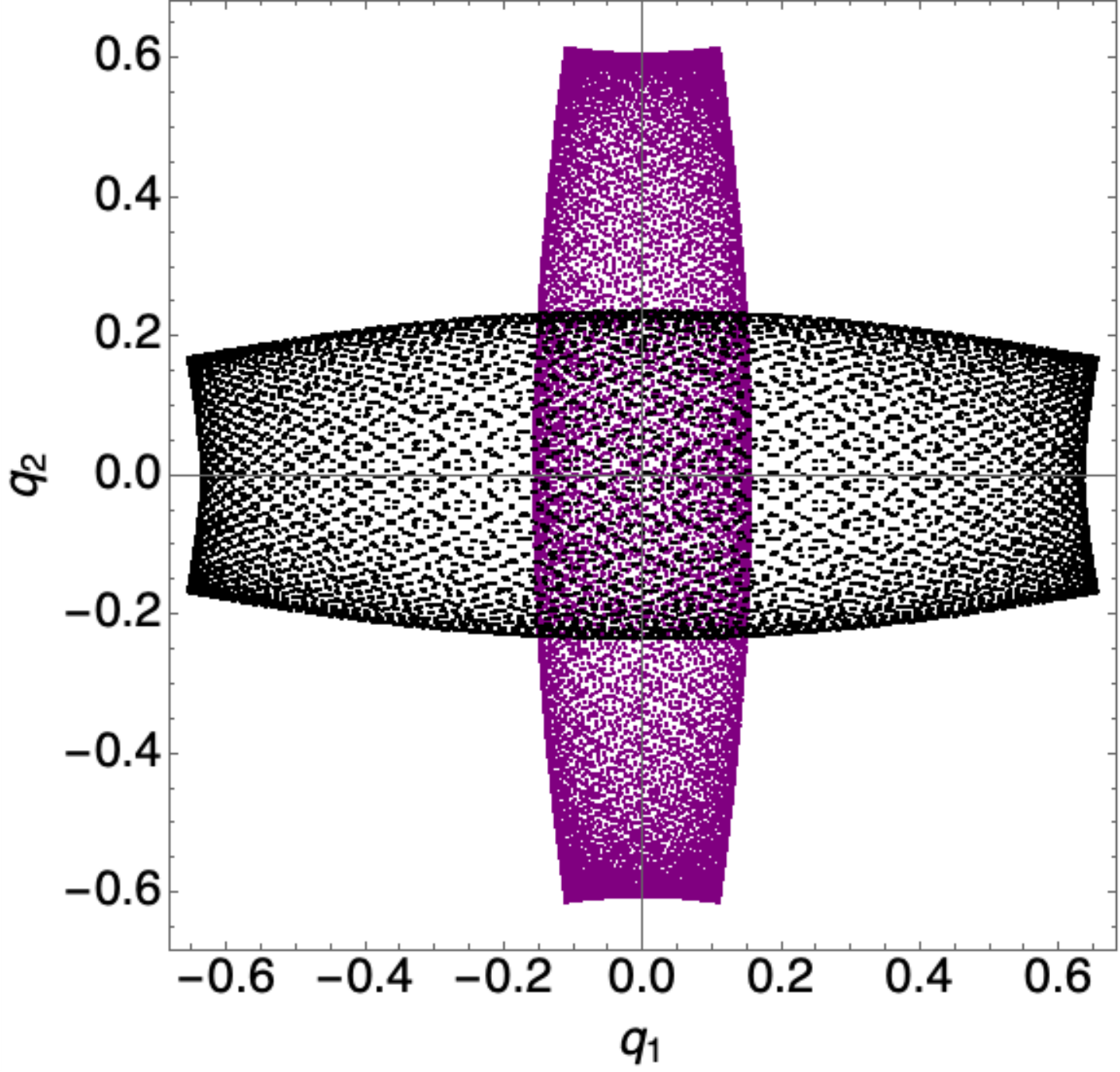}
\includegraphics[scale=0.29]{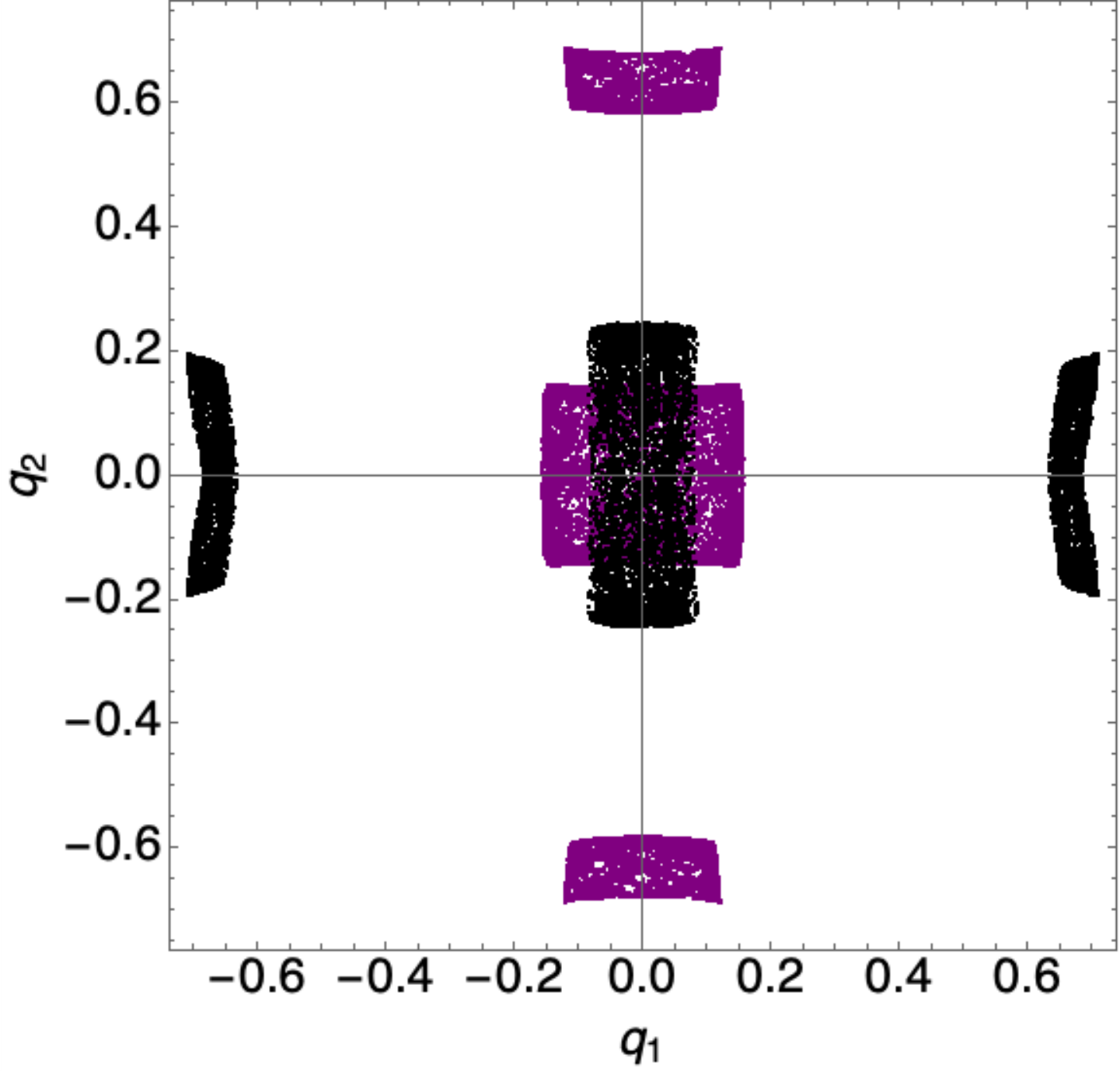}
\includegraphics[scale=0.2]{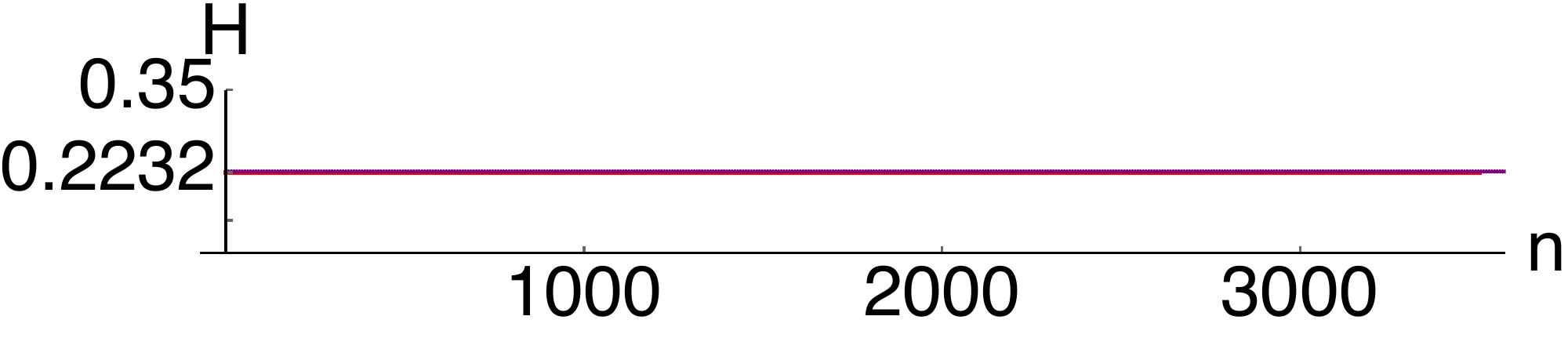}
\includegraphics[scale=0.2]{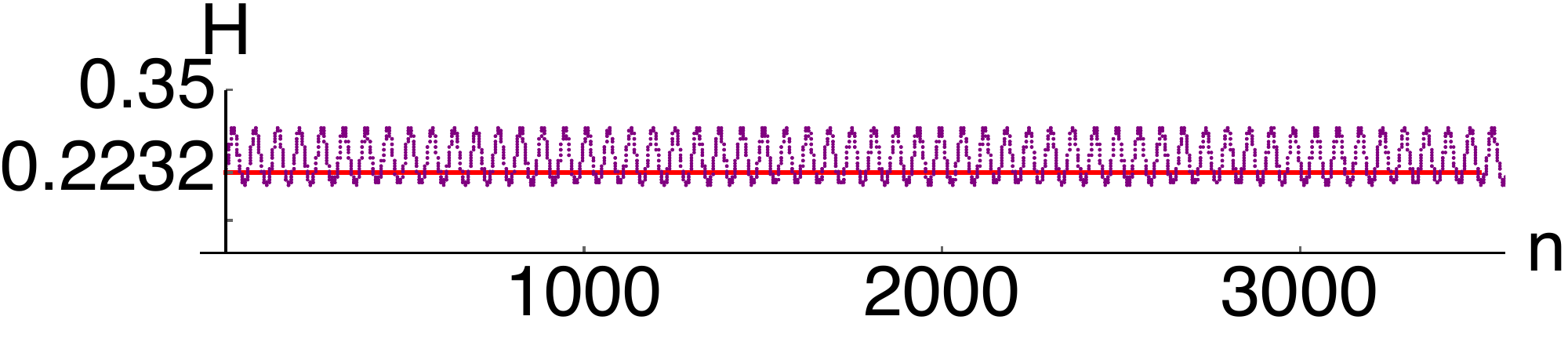}
\includegraphics[scale=0.2]{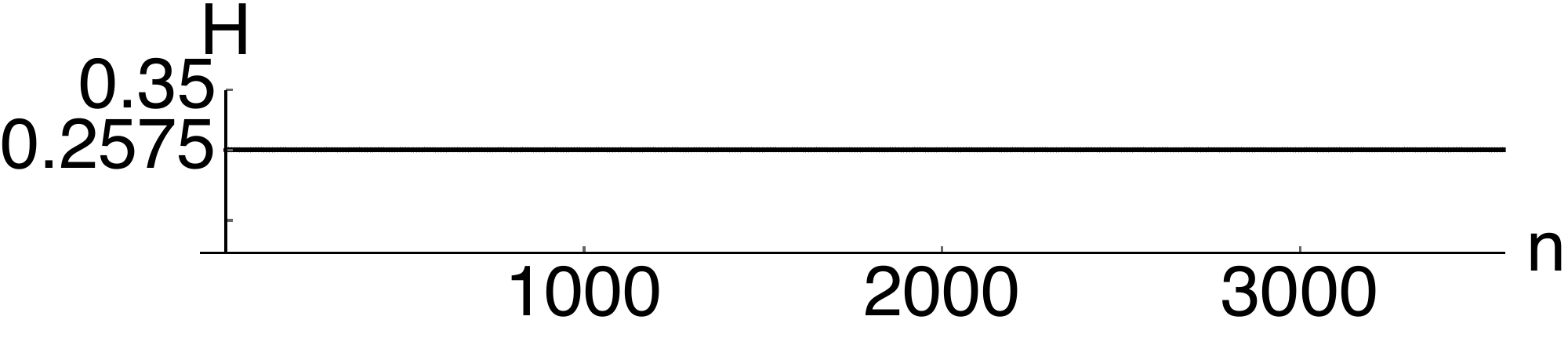}
\includegraphics[scale=0.2]{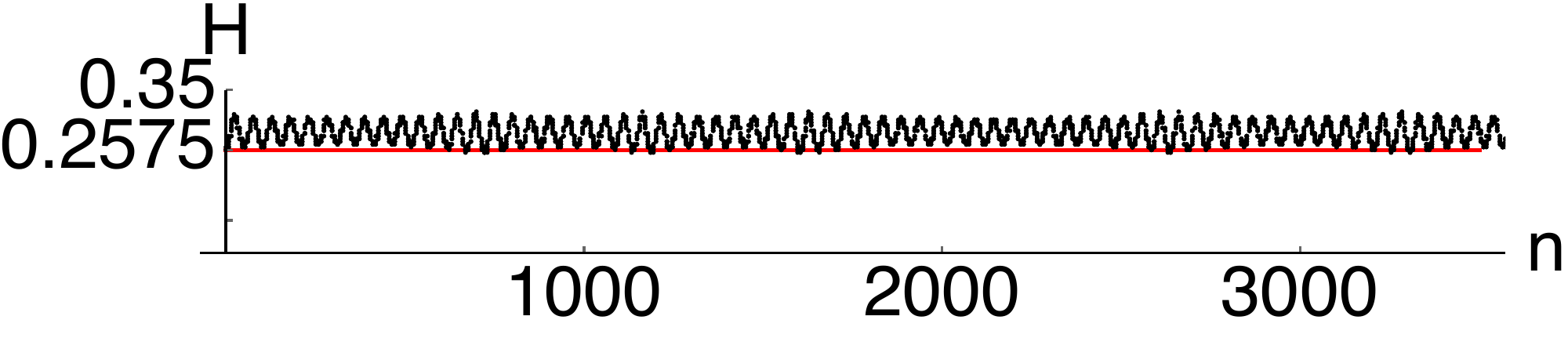}\\
\caption{Projection of two trajectories on the $(q_1,q_2)$ plane for the equidistant lattice with the octupole magnet as a nonlinear element (upper right panel) and Ruth lattice with the octupole magnet as a nonlinear element (upper left panel). Hamiltonian  Eq.\eqref{eq:HH2} as a function of the iteration number $n$ for the equidistant lattice (lower right panels) and Ruth lattice (lower left panels). Black line and purple line - is the tracking and red line is the exact value of the Hamiltonian Eq.\eqref{eq:HH2}. Purple corresponds to the initial conditions $q_1=-0.11$, $p_1=0$, $q_2=0.6$, $p_2=0.15$ and black corresponds to the initial conditions $q_1=0.64$, $p_1=0$, $q_2=0.063$, $p_2=0.15$.}
\label{Fig:8}
\end {figure}

 \begin {figure*}
 \centering
\includegraphics[scale=0.4]{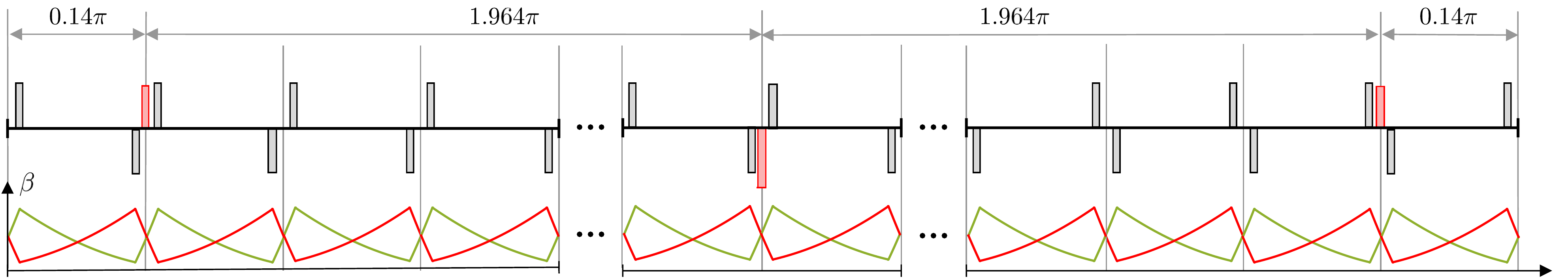}
\caption{Schematics of the Yoshida lattice layout (upper panel) and corresponding $\beta$-functions plot (lower panel).}
\label{Fig:9}
\end {figure*}    

As before, we build a Poincare surface of section $q_2=0$ for the equidistant lattice and Ruth lattice by tracking several sets of initial conditions for $10^6$ iterations and compare it to the Poincare surface of section of a smooth Hamiltonian Eq.\eqref{eq:HH2} in Fig.\ref{Fig:7e} and Fig.\ref{Fig:7} correspondently. We utilized the same technique as in Sec.\ref{sec:rsex} to produce these surface of sections.  For the comparison we considered level set $\mathrm{H}=0.2$ with $\mathrm{H}$ defined by Eq.\eqref{eq:HH2}. We picked the tune of the linear lattice $\nu=0.2234$ slightly below the characteristic resonance $1/4$ of the octupole. By comparing upper left and upper right plots in Fig.\ref{Fig:7} we again observe that the topology of both surfaces of section coincide quite well and the Hamiltonian Eq.\eqref{eq:HH2} (Fig.\ref{Fig:7} bottom panel) is conserved with high accuracy, as expected. Tracking of the equidistant lattice with the potential scaling given by Eq.\eqref{eq:DNpO} again  failed as part of the considered initial conditions rapidly escaped to infinity as illustrated in Fig.\ref{Fig:7e}.

On can notice that as in the case of a Ruth lattice with the sextupoles Poincare surfaces of section for a smooth system and for the Ruth lattice with octupoles do not coincide completely. This again could be explained by a high value of the parameter $h\approx0.28$ - equivalent integrator step in phase. 

In order to highlight the significant difference between the equidistant lattice and Ruth lattice we track two sets of initial conditions near the separatrix $\mathrm{H}=1/4$. In Fig.\ref{Fig:8}, purple color corresponds to the level set $\mathrm{H}=0.2232$ slightly below the separatrix and the black color to the level set $\mathrm{H}=0.2575$ - slightly above the separatrix. 
From Fig.\ref{Fig:8} we conclude, that the motion is quite different and the Hamiltonian in case of the equidistant lattice is conserved poorly. We also note that for the equidistant lattice, projection of the trajectories splits into regions and form an island-like structure. It is worth mentioning that recently such structure of the projection was observed experimentally at IOTA \cite{NK} with the help of the SyncLight system \cite{NKs}, though the number of nonlinear magnets used in the experiment was 17 and not 5 as in the present numerical example.

\subsection{Yoshida lattice} \label{sec:yosh} 
  
A linear lattice that mimics the Yoshida integration scheme should correspond to the modified integrator structure given by Eq.\eqref{eq:Yoshf}. It is schematically depicted in Fig.\ref{Fig:3}(c). As seen from Eq.\eqref{eq:Yoshf}, only three nonlinear elements are needed, and the physical distances between them could be quite large. Locations of the nonlinear elements are defined by the integrator coefficients, $\gamma_1$ and $\kappa_1$, and relative strength of the magnetic field by $\gamma_1$ and $\kappa_2$. Positions of the magnets also depend on the choice of $h$ - the integrator step in phase that could be an arbitrary number less then unity. The total phase advance of the linear lattice should be $4\pi+h$ as follows again from Eq.\eqref{eq:Yoshf}.

To illustrate the Yoshida lattice we picked the simplest example where linear optics is implemented with FODO cells without bends. For this example we treat quadrupoles as infinitely thin lenses. The phase advance of one cell is chosen as prescribed by the first rotation in Eq.\eqref{eq:Yoshf} - $h\gamma_1/2$         

For the specific example, linear optics is implemented with 30 FODO cells (see Fig.\ref{Fig:9}) with the phase advance of one cell equal to $0.14\pi$. The total phase advance of the channel is $4.208\pi$ and corresponding step of the integrator \eqref{eq:Yoshf} is $h=0.208\pi$.      

It was suggested in Ref.\cite{DanNag} to design a machine with the following effective Hamiltonian

\begin{align} 
\label{eq:NLPh}
\mathrm{H}=\frac{p_1^2+p_2^2}{2}+\frac{q_1^2+q_2^2}{2}+V_{\mathrm{Dr}}(q_1,q_2).
\end{align}  
here  $V_{\mathrm{Dr}}(q_1,q_2)$ is the Darboux potential (a solution to the Bertrand-Darboux partial differential equation) given by
\begin{align}
\label{eq:NLP}
V_{\mathrm{Dr}}(q_1,q_2)&=\frac{f_2\left[ \xi(q_1,q_2)\right]+g_2\left[\eta(q_1,q_2)\right]}{\xi(q_1,q_2)^2-\eta(q_1,q_2)^2}, \\
f_2(\xi)&=-0.4\mathrm{arccosh}(\xi ) \xi  \sqrt{\xi ^2-1} , \nonumber \\
g_2(\eta)&=-0.4\left[\mathrm{arccos}(\eta )-\frac{\pi }{2}\right]\eta  \sqrt{1-\eta ^2},  \nonumber\\
\xi(q_1,q_2)&=\frac{\sqrt{\left(q_1+1\right){}^2+q_2^2}+\sqrt{\left(q_1-1\right){}^2+q_2^2}}{2} ,\nonumber\\
\eta(q_1,q_2)&=\frac{
   \sqrt{\left(q_1+1\right){}^2+q_2^2}-\sqrt{\left(q_1-1\right){}^2+q_2^2}}{2} \nonumber.
\end{align}
More details on this potential and corresponding nonlinear magnet could be found in the recent publication by Mitchell in Ref.\cite{chad_r}.
 
Hamiltonian Eq.\eqref{eq:NLPh} is completely integrable and thus has infinite region of stable and regular particle motion (infinite dynamic aperture). The first integral of motion is the Hamiltonian itself and second integral of motion is given by \cite{DanNag,Bella}
\begin{align}
\label{eq:SIM}
I_2=p_1^2+(p_2q_1-p_1q_2)^2+2\frac{\eta^2f(\xi)+\xi^2g(\eta)}{\xi^2-\eta^2}.
\end{align} 
With $\xi(q_1,q_2)$ and $\eta(q_1,q_2)$ given as in Eq.\eqref{eq:NLP} and 
\begin{align}
f(\xi)&=\frac{\xi^2(\xi^2-1)}{2}+f_2(\xi)  \nonumber,\\
g(\eta)&=\frac{\eta^2(1-\eta^2)}{2}+g_2(\eta) .
\end{align}
\begin {figure}[t]
 \centering
\includegraphics[scale=0.209]{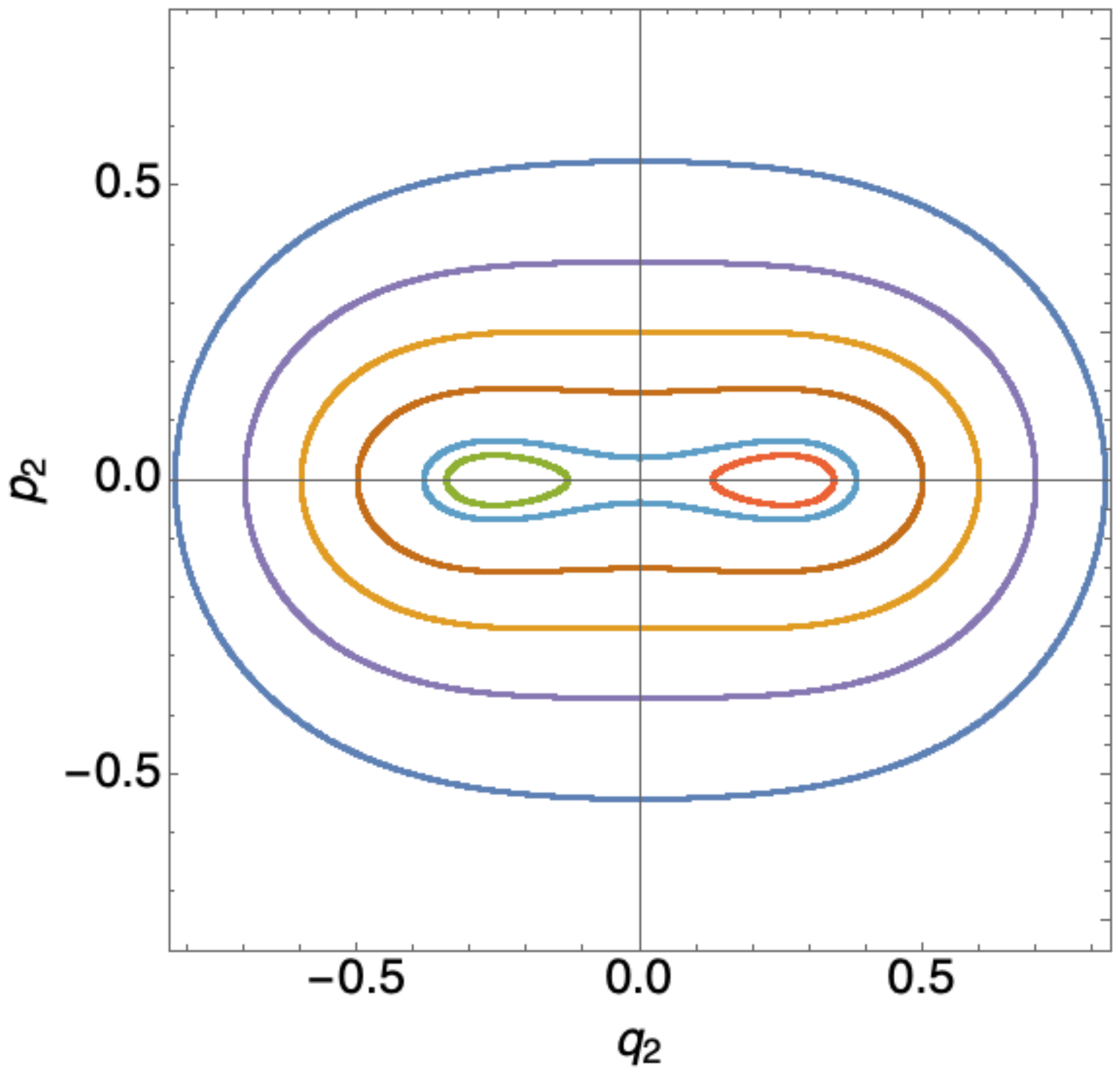}
\includegraphics[scale=0.209]{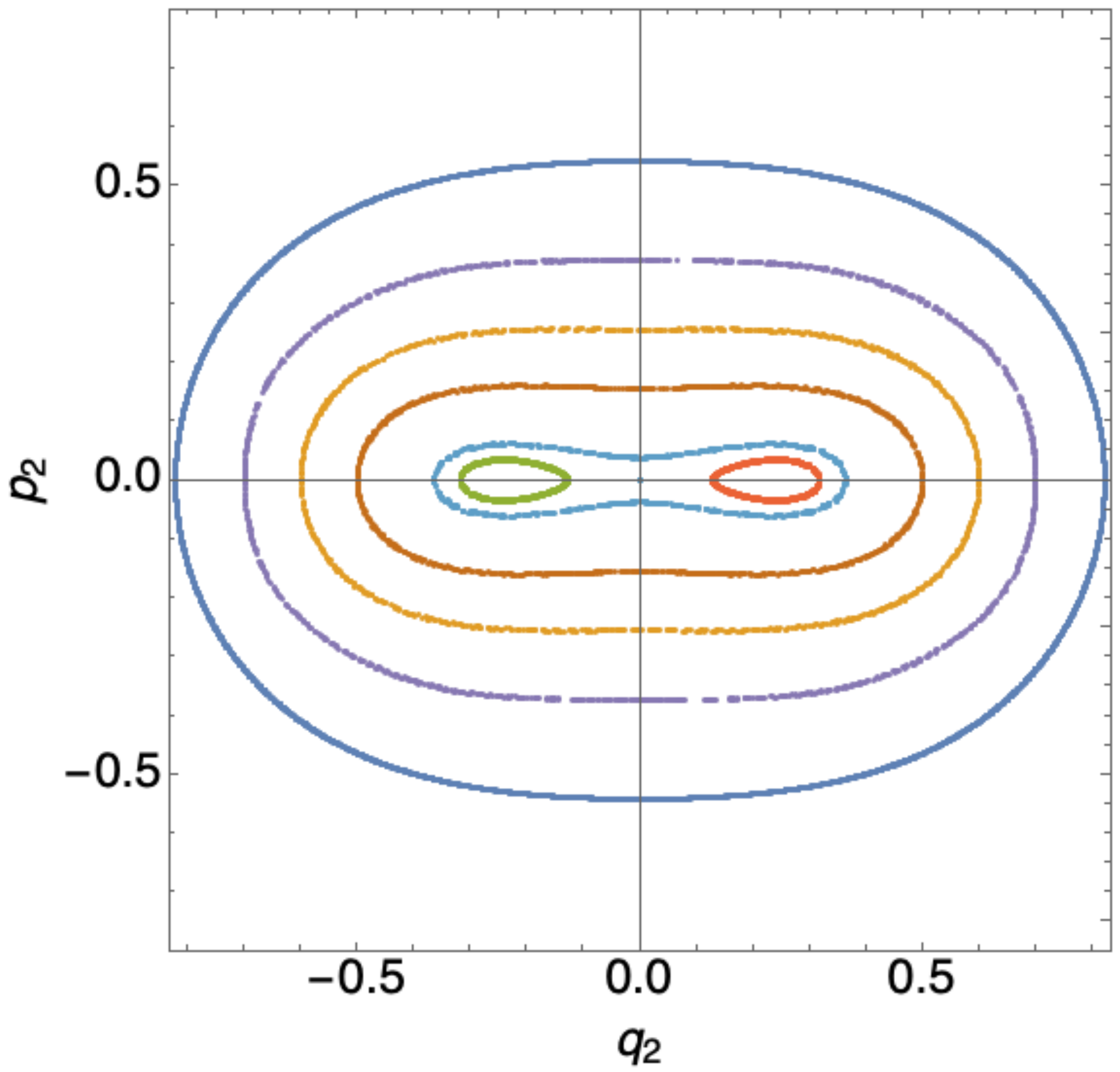} 
\includegraphics[scale=0.43]{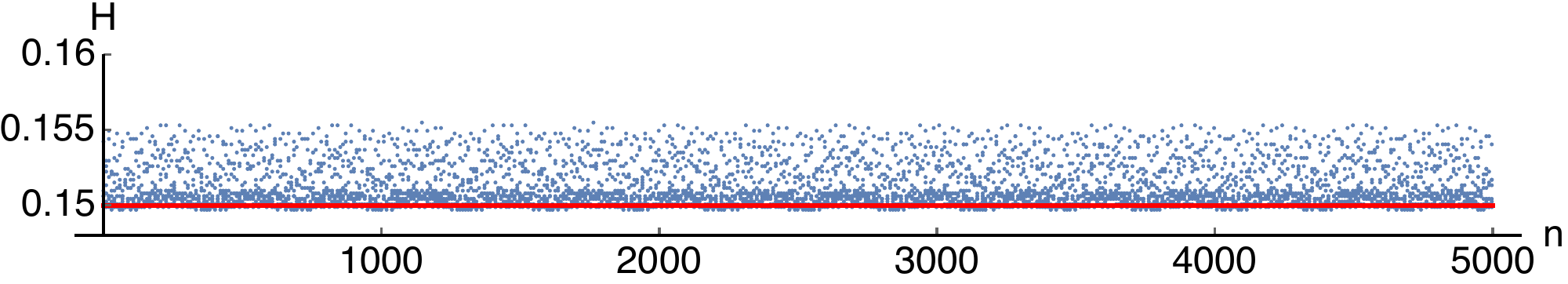}\\
\includegraphics[scale=0.43]{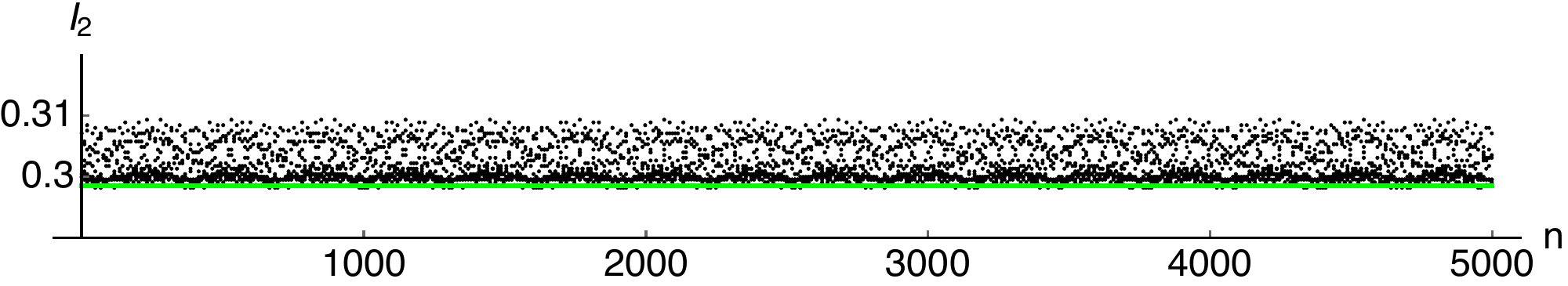}\\
\caption{Poincare surface of section $q_1=0$ for the smooth Hamiltonian \eqref{eq:NLPh} with the Darboux potential \eqref{eq:NLP} (upper left panel) and Yoshida lattice shown on Fig.\ref{Fig:10} tracked for $5\times 10^5$ iterations (upper right plane). Hamiltonian Eq.\eqref{eq:NLPh} and second integral of motion Eq.\eqref{eq:SIM} (lower panels) lines (red and green) are for the smooth system, dots (blue and black) are for the lattice tracking.}
\label{Fig:10}
\end {figure}        

We have considered thin nonlinear lens with the potential given by Eq.\eqref{eq:NLP} as a nonlinear element for the Yoshida lattice (red blocks in Fig.\ref{Fig:9}) and calculated corresponding transformations of these lenses according to the prescription of Eq.\eqref{eq:nlt} and Eq.\eqref{eq:nlt2}. Exact values of the $\beta$-function as well as the $\beta$-plot in Fig.\ref{Fig:9} were calculated using ``OptiMX" software \cite{optim}. 

We build a Poincare surface of section $q_1=0$ for the Yoshida lattice by tracking several sets of initial conditions for $5\times10^5$ iterations and compare it to the Poincare surface of section of a smooth Hamiltonian given by Eq.\eqref{eq:NLPh}. In Fig.\ref{Fig:10} we show results for both the tracking of Yoshida lattice and a smooth system. For the comparison we took the level set $\mathrm{H}=0.15$.  The same technique as described in Sec.\ref{sec:rsex} was utilized to produce these plots.   

As expected we see that the topology of the Poincare surface of section is the same, and moreover it looks like they are almost identical. We also observe that the Hamiltonian Eq.\eqref{eq:NLPh} as well as the second integral of motion Eq.\eqref{eq:SIM} are conserved within $\approx3\%$ accuracy. 

Less amount of points on the right top plot Fig.\ref{Fig:10} in comparison to the left top plot  Fig.\ref{Fig:10} is explained by a difference in methods when constructing the Poincare surface of section as well as relatively low number of iterations for the Yoshida lattice tracking.  

This illustration explicitly demonstrates that the nonlinear integrable channel introduced in Ref.\cite{DanNag} could be implemented with as low as just three nonlinear elements with the field that corresponds to the Darboux potential given by Eq.\eqref{eq:NLP}. It is worth mentioning that nonlinear magnets in the Yoshida lattice could be any nonlinear lens, be it sextupole, octupole or any other potential that could be implemented using magnetic coils. 
    
\section{Conclusion}\label{sec:con}

We have presented a new approach to the nonlinear optics design strategy based on  symplectic integration schemes. We demonstrated how one may utilize known symplectic integrators to produce optics configurations that preserve given Hamiltonians with any desired degree of accuracy. As relevant examples, we considered the Ruth integrator of the second order and the Yoshida integrator of the forth order to produce optics lattices.

In the presented examples we have demonstrated that the current design of the octupole and nonlinear channels under active study at the IOTA facility could be improved with just a few modifications. We showed as well that with just three nonlinear magnets one can produce a lattice that will still conserve the given Hamiltonian and thus have a large region of stable and regular particle motion. 

We would like to highlight that the suggested method of lattice design is independent of the choice of nonlinear potential and thus could be applied for any nonlinear lens, be it conventional sextupole and octupole lens, or nonlinear lens with the Darboux potential \cite{DanNag}. It may even be expanded to the case of electron lenses \cite{el1,el2} with some modifications.

The approach that we introduced may also serve as an initial seed for the conventional lattice design and could be incorporated into the lattice optimization workflow.      

\appendix 

\section{Recovery of the Hamiltonian from the Ref.\cite{DanNag}\label{app:DNlim}}

In this section we recover original scaling of the nonlinear potential with the $\beta$-function from the Ref.\cite{DanNag}.
First we observe that if $h\to0$ and $\beta_x=\beta_y=\beta$ then substitution 
\begin{align}
h=\frac{ds}{\beta(s)} 
\end{align}
is exact and the flow that is given by Eq.\eqref{eq:nlt} could be rewritten as  
\begin{align}
\label{eq:nltm}
N^{\beta}\mathrm{X}_0=\left[x^0,P^0_x-\frac{ds}{\beta(s)} \frac{\partial_{x}U}{\sqrt{\beta(s_i)}},y^0,P_y^{0}-\frac{ds}{\beta(s)}\frac{\partial_{y}U}{\sqrt{\beta(s_i)}}\right],
\end{align}
with the additional substitution for the $\partial_{x,y}U$ defined in Eq.\eqref{eq:nlt2}.

It is apparent that Hamiltonian that corresponds to this flow \eqref{eq:nltm} is exactly 
\begin{align}
\label{eq:DNpAp}
U(x,y,s)=\frac{1}{\beta(s)}V\left(\frac{x}{\sqrt{\beta(s)}},\frac{y}{\sqrt{\beta(s)}}\right).  
\end{align} 
As far as the flow that is given by Eq.\eqref{eq:ltr} is generated by a Hamiltonian 
\begin{align}
\mathrm{H}=&\frac{P_x^2+P_y^2}{2}+K(s)\left(\frac{x^2+y^2}{2} \right)
\end{align}
the final Hamiltonian that generates the flow 
\begin{align}
N^{\beta}\circ M_{x,y}(s_2|s_1)
\end{align} 
in the case of $h\to 0$ is exactly the one from the Ref.\cite{DanNag} and is given by
\begin{align}
\mathrm{H}=&\frac{P_x^2+P_y^2}{2}+K(s)\left(\frac{x^2+y^2}{2} \right)\nonumber\\ &+\frac{1}{\beta(s)}V\left(\frac{x}{\sqrt{\beta(s)}},\frac{y}{\sqrt{\beta(s)}}\right).
\end{align}
It is worth to mention that differences in scaling coefficient that were mentioned in Sec.\ref{sec:rsex} and  Sec.\ref{sec:roct} vanish in the limit of $h\to0$. Indeed, as mentioned above, substitution $h=ds/\beta$ is exact in this case and thus there is no difference in using $h$ or $ds$ in therms of an integrator. However, switching to $ds$ introduces additional factor $1/\beta$ as one may notice from Eq.\eqref{eq:nltm}.

\section{Cross-check with the MADX\label{app:madx}}

To produce such a comparison we consider Yoshida lattices that is described in Sec.\ref{sec:examp} and schematically presented in Fig.\ref{Fig:9} -  30 FODO cells with the phase advance of one cell equal to $0.14\pi$. The total phase advance of the channel is $4.208\pi$ and corresponding step of the integrator \eqref{eq:Yoshf} is $h = 0.208\pi$.  
A sextupole was picked as a nonlinear element instead of a magnet with the Darboux potential. Consequently, the invariant of motion that this lattice is aimed to conserve is a Henon-Heiles Hamiltonian in normalized coordinates that is given by Eq.\eqref{eq:HH}.

\begin {figure}[t]
 \centering
\includegraphics[scale=0.209]{Fig5a.pdf}
\includegraphics[scale=0.209]{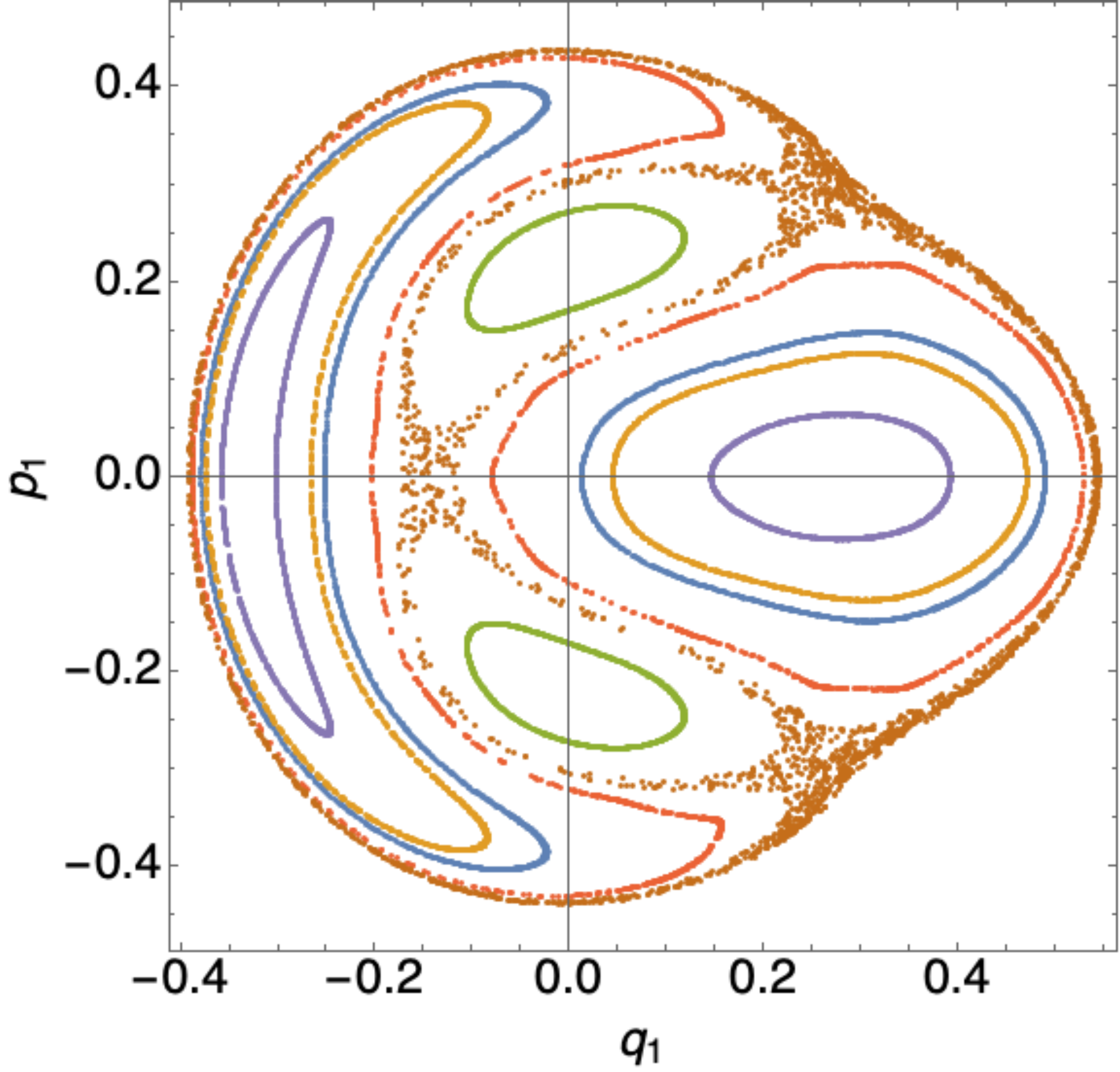}
\includegraphics[scale=0.43]{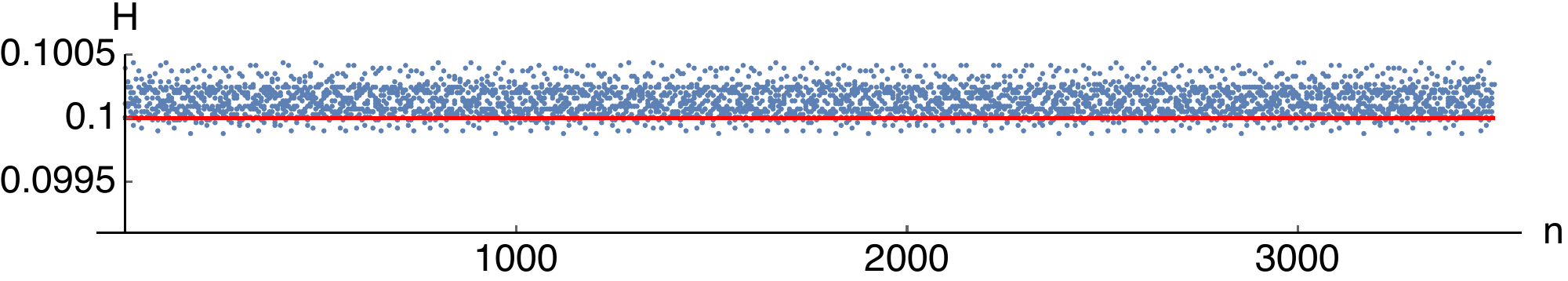}\\
\caption{Poincare surface of section $q_2=0$ for the smooth Henon-Heiles Hamiltonian Eq.\eqref{eq:HH} (left panel) and MADX tracking of the Yoshida lattice with the sextupole magnet as a nonlinear element (right panel). Hamiltonian  Eq.\eqref{eq:HH} as a function of the iteration number $n$ (lower panel). Blue line - MADX tracking for the Yoshida lattice for the $10^6$ turns and red line is the exact Hamiltonian of the smooth system.}
\label{Fig:11}
\end {figure}

In Fig.\ref{Fig:11} we present Poincare surface of sections ($q_2=0$) for the smooth Hamiltonian Eq.\eqref{eq:HH} and Yoshida lattice with sextupoles. We observe, that as before lattice indeed conserves given Hamiltonian as prescribed. For the MADX input initial conditions that are listed in Table \ref{tab:Table1} were transformed to the real coordinate space with the help of the betatron amplitude matrix given by Eq.\eqref{eq:BTM} and linear lattice functions that were calculated initially using OptiMX \cite{optim} and then confirmed by MADX.

MADX input file as well as Wolfram Mathematica post-processing script could be found in supplemental materials.

\section{Definition of some terms used in the paper\label{app:df}}

\textit{Bounded Motion} - the motion of is called bounded if for the initial conditions $X_0$ and a state vector $X$ there exist some constant $C$ such that $|X|\leq C$. 

{\textit{Dynamic Aperture} - maximum phase space volume that contains bounded trajectories of the system.

\textit{Chaotic motion} - a random motion of a deterministic system.    

\textit{Integrable system} - a Hamiltonian system with $2n$ dimensional phase space that has the maximal number of independent Poisson commuting invariants (including the Hamiltonian itself) equals to $n$.
 
\textit{Quasi-integrable system} - a Hamiltonian system that has approximate Poisson commuting invariants that are close to the invariants of some integrable system.

\textit{Smooth Hamiltonian} - a Hamiltonian of a system with continuous time.

\textit{Effective Hamiltonian} - a smooth Hamiltonian that captures the dynamics of a discreet system.

\section{Parameters for the tracking}
 
 In this Appendix in order  for the reader to be able to reproduce tracking results we present sets of initial conditions that were usec to produce Fig.\ref{Fig:5e},Fig.\ref{Fig:5}, Fig.\ref{Fig:7e}, Fig.\ref{Fig:7} and Fig.\ref{Fig:10}. Note that coordinates are given in normalized (canonical) coordinates that are connected to the real coordinates through the betatron amplitude matrix \cite{SYL}.
 
 \begin {table}[h!]
\caption{Initial conditions for the Poincare surface of section Fig.\ref{Fig:5} for the sextupole lattice and Henon-Heiles Hamiltonian Eq.\eqref{eq:HH}.}
\label{tab:Table1}
\begin{ruledtabular}
\begin {tabular}{c c c c c}
set number&$q_1$&$p_1$& $q_2$&$p_2$ \\
\colrule
1&0.095 & 0.096& 0&0.427003 \\
2&0.15&0.0960&0&0.412958 \\
3&-0.1&0.157&0&0.405814\\
4&-0.2&0.05&0&0.390086\\
5&0.2&0.05&0&0.403527\\
6&-0.12&0.005&0&0.429445
\end {tabular}
\end{ruledtabular}
\end{table}

 \begin {table}[h!]
\caption{Initial conditions for the Poincare surface of section Fig.\ref{Fig:7} for the octupole lattice and Henon-Heiles Hamiltonian Eq.\eqref{eq:HH2}.}
\label{tab:Table2}
\begin{ruledtabular}
\begin {tabular}{c c c c c}
set number&$q_1$&$p_1$& $q_2$&$p_2$ \\
\colrule
1&0.02&0.06&0&0.629285 \\
2&0.04&0.12&0&0.619676 \\
3&0.06&0.18&0&0.603319\\
4&0.078&0.22&0&0.58779\\
5&0.1&0.3&0&0.547677\\
6&0.12&0.36&0&0.505862\\
7&0.14&0.42&0&0.451451\\
8&0.16&0.48&0&0.379041\\
9&0.18&0.54&0&0.274727\\
10&0.4&0.05&0&0.474025
\end {tabular}
\end{ruledtabular}
\end{table}
   
 \begin {table}[h!]
\caption{Initial conditions for the Poincare surface of section Fig.\ref{Fig:10} for the Yoshida lattice composed of three nonlinear magnets with the potential given by Eq.\eqref{eq:NLP}.}
\label{tab:Table3}
\begin{ruledtabular}
\begin {tabular}{c c c c c}
set number&$q_1$&$p_1$& $q_2$&$p_2$ \\
\colrule
1&0&0.07967&0.1& 0.54 \\
2&0&0.41716&0.6&0.01 \\
3&0&0.54447&0.13&0.005 \\
4&0&0.54447&-0.13&-0.005 \\
5&0&0.33064&0.7&0.01 \\
6&0&0.47124&0.5&0.01 \\
7&0&0.54597&0.04&0.04
\end {tabular}
\end{ruledtabular}
\end{table}   

\newpage

\begin{acknowledgments}
This work was supported by the U.S. National Science Foundation under Award No. PHY-1549132, the Center for Bright Beams and under Award No. PHY-1535639;
The author is grateful to Sergei Nagaitsev for the introduction to the problem and several fruitful discussions, to Alexander Valishev for couple of useful remarks, to Ivan Bazarov for sharing Ref.\cite{GInt}, to Gerard Andonian for reading the manuscript.    
\end{acknowledgments}

\bibliographystyle{ieeetr}
\bibliography{Integrator}

\end{document}